\documentclass[pra,aps,twocolumn,showpacs,superscriptaddress,nofootinbib,floatfix]{revtex4-1}
\usepackage[colorlinks=true,citecolor=blue,linkcolor=magenta]{hyperref}
\usepackage{graphicx,stackrel}
\usepackage{dcolumn}
\usepackage{bm}
\usepackage[bbgreekl]{mathbbol}
\usepackage{amsmath,dsfont}
\usepackage[normalem]{ulem}
\usepackage{color}
\usepackage[ruled,vlined]{algorithm2e}

\usepackage{booktabs}
\def\ii{{\rm i}} 
\newcommand{\Ham}{\hat{\mathcal{H}}}
\newcommand{\Obs}{\hat{\mathcal{O}}}
\newcommand{\Graph}{{\mathcal{G}}}
\newcommand{\edges}{{\mathcal{E}}}
\newcommand{\vertices}{{\mathcal{V}}}
\newcommand{\e}{\mathbf{\text{e}}}
\newcommand{\Sum}{\displaystyle\sum}
\newcommand{\Prod}{\displaystyle\prod}
\newcommand{\Int}{\displaystyle\int}
\newcommand{\Sin}{\sin\vartheta}
\newcommand{\Cos}{\cos\vartheta}
\newcommand{\Tan}{\tan\vartheta}
\newcommand{\Id}{\mathbb{1}}
\def\braket#1{\mathinner{\langle{#1}\rangle}}
\newcommand{\bra}[1]{\left\langle #1\right|}
\newcommand{\ket}[1]{\left|#1\right\rangle}

\def\Real#1{\mathinner{\Re\left\{#1\right\}}}
\newcommand{\alphab}{\boldsymbol{\alpha}}

\begin{document}
\title{Quantum evolution kernel : Machine learning on graphs with programmable arrays of qubits}
\author{Louis-Paul Henry}
\thanks{These authors contributed equally to this work}
\affiliation{Pasqal, 2 avenue Augustin Fresnel, 91120 Palaiseau}

\author{Slimane Thabet}
\thanks{These authors contributed equally to this work}
\affiliation{Pasqal, 2 avenue Augustin Fresnel, 91120 Palaiseau}

\author{Constantin Dalyac}
\affiliation{Pasqal, 2 avenue Augustin Fresnel, 91120 Palaiseau}
\affiliation{LIP6, CNRS, Sorbonne Université, 4 Place Jussieu, 75005 Paris, France}

\author{Loïc Henriet}
\email{loic@pasqal.io}
\affiliation{Pasqal, 2 avenue Augustin Fresnel, 91120 Palaiseau}

\date{\today}
\begin{abstract}
The rapid development of reliable Quantum Processing Units (QPU) opens up novel computational opportunities for machine learning. Here, we introduce a procedure for measuring the similarity between graph-structured data, based on the time-evolution of a quantum system. By encoding the topology of the input graph in 
the Hamiltonian of the system, the evolution produces measurement samples that retain key features of the data. We study analytically the procedure and illustrate its versatility in providing links to standard classical approaches. We then show numerically that this scheme performs well compared to standard graph kernels on typical benchmark datasets. Finally, we study the possibility of a concrete implementation on a realistic neutral-atom quantum processor. 
\end{abstract}
\maketitle

\section*{Introduction}

Machine learning has now become part of the arsenal of almost
all scientific fields.  It has been successfully used in
a wide range of applications, in particular whenever data can be represented as vectors (of potentially very high dimension). This is especially relevant in the analysis of image data and natural language processing.

In many fields such as chemistry \cite{Varnek2012,gilmer17}, bioinformatics \cite{Muzio20,Borgwardt05}, 
social network analysis\cite{Scott11}, computer vision\cite{harchaoui2007image}, or natural language processing \cite{nikolentzos2017shortest, glavavs2013recognizing}, some observations have an inherent graph structure,
requiring the development of specific algorithms to exploit the information contained in their structures. 
A large body of work exists in the classical machine learning literature, trying to study graphs through the use of {\sl graph kernels} that are measures of similarity between graphs\cite{Latouche15,Kriege20,Borgwardt20,Schlkopf01}. 
The richness of graphs' structure gives them their versatility but poses a challenge 
in how one ought to proceed in order to extract the information they contain.
The idea behind the graph kernel approach is very
generic, and consists first in finding a way to associate any graph with a {\sl feature vector} encapsulating its relevant characteristics (the {\it feature map}) and then to compute the similarity between those vectors, in the form of a scalar product in the feature space.  The design of a kernel always comes down to a trade-off between capturing enough characteristics of 
the graph structure and not being too specific (so that graphs with similar properties yield similar results) while still being algorithmically efficient.

A lot of kernels have been introduced, each particularly fit to the study of particular types of problems or graph structures. 
The algorithmic complexity of most of these kernels makes them difficult to apply to the study of large graphs. 
For example, the {\sl graphlets} subsampling and the random walk kernels that we will be using later scale respectively as $N^k$ and $N^3$ with the graph size $N$, and $k$ the graphlet size (see \ref{sec:appendix classical kernels}).

Previous work on quantum graph kernels has been done\cite{Schuld19,Havlek19,kishi21}, both as generic algorithms
introducing novel approaches inspired by quantum mechanics (Quantum graph neural networks\cite{Verdon19}, Quantum-walk kernels\cite{Rossi15,Bai15}), as well as with 
a more specific computing platform in mind ({\sl e.g.} using photonic devices\cite{Schuld20}).

Here we propose a more versatile and easily scalable graph kernel based on similar ideas. The core principle is to encode the information about a graph in the parameters of a tunable Hamiltonian acting on an array of qubits 
and to measure a carefully chosen observable after an alternating sequence of {\it free} evolution ({\sl i.e.} with this Hamiltonian) and/or parametrized 
pulses, similarly to what is done in the Quantum Approximate Optimization Algorithm (QAOA)\cite{Farhi14}, or after a continuously parametrization of the Hamiltonian, similarly to what can be done in optimal control\cite{Werschnik07}.
This kernel can be realized with state-of-the-art QPUs, in particular with Rydberg atom processors \cite{Weimer10,Bernien17,Labuhn16}, in which highly tunable Hamiltonians can be realized to encode a wide range of graph topologies with up to hundreds of qubits.

The paper is structured as follows. 
We first briefly introduce the reader to graph kernels in \ref{sec:GraphKernels}.
We then present the generic ideas of our quantum kernel in section \ref{sec:QuantumGraphKernel}. 
An example of this kernel is detailed in \ref{sec:Ising} illustrating how it probes the underlying graph structure, and its
versatility is illustrated in \ref{sec:ExistingKernels} where it is related to other existing graph kernels.
The performances of this kernel on real datasets are compared to other existing kernels in \ref{sec:Benchmarking} and its implementation on neutral-atom processors is discussed and characterized in \ref{sec:implementation}.


\section{ Graph kernels}
\label{sec:GraphKernels}
One approach allowing to apply the tools of machine learning to data in the form of graphs is to find a way to represent each graph in an {\it embedding } vector space, and then measure the similarity between two graphs as the inner product between their representative vectors.

A graph kernel $K$ constitutes such an inner product and therefore allows to perform machine learning tasks on graphs. 
More specifically, a graph kernel is a symmetric,  positive semidefinite function defined on the space of graphs $\mathbb{G}$.  
Given a kernel $K$, there exists a map 
$\phi:\mathbb{G} \to \mathcal{H}$ into a Hilbert space $\mathcal{H}$ such that $K(\Graph_1,\Graph_2)  =\braket{\phi(\Graph_1)|\phi(\Graph_2)}$
for  all $\Graph_1,\Graph_2 \in \mathbb{G}$\cite{nikolentzos19}.

For completeness, we detail here two algorithms in which graph kernels can be used: Support Vector Machine (SVM) for classification 
({\sl i.e.} sorting data in categories) and Kernel Ridge Regression (KRR) for regression ({\sl i.e.} the prediction of continuous values)\cite{Schlkopf01,Bishop06}.
These methods have been successfully applied to data sets of graphs of up to a few dozen nodes \cite{Morris+2020}.

\subsection{Support Vector Machine\label{sec:SVM}}
The SVM algorithm aims at splitting a dataset in two classes by finding the best hyperplane that separates the data points in the feature space, in which the coordinates of each data point (here each graph) is determined according to the kernel $K$.

For a training graph dataset $\left\{\Graph_i\right\}_{i=1\ldots M}$, and a set of labels ${\bf y}=\left\{y_i\right\}_{i=1\ldots M}$ 
(where $y_i=\pm1$ depending on which class the graph $\Graph_i$ belongs to), 
the dual formulation of the SVM problem consists in finding 
$\alphab_0\in\mathcal{A}_C({\bf y})=\left\{\alphab\in[0,C]^{M}\right|\alphab^T{\bf y} = 0\}$
such that
\begin{gather}
    \frac{1}{2}\alphab_0^TQ\alphab_0 - {\bf e}^T\alphab_0
    =\min_{\alphab\in\mathcal{A}_C({\bf y})}\;\left\{\frac{1}{2}\alphab^TQ\alphab - {\bf e}^T\alphab\right\}
\end{gather}
where ${\bf e}$ is the vector of all ones, $Q$ is a $M\times M$ matrix such that $Q_{ij} = y_i y_j K(\Graph_i,\Graph_j)$, and $C$ is the penalty hyperparameter, to be adjusted. Setting $C$ to a large value increases the range of possible values of $\alpha$ and therefore the flexibility of the model. But it also increases the training time and the risk of overfitting.

The data points for which $\alpha_i>0$ are called support vectors (SV). Once the $\alpha_i$ are trained, the class of a new
graph $\Graph$ is predicted by the decision function, given by:
\begin{align}
    \label{eq:SVM}
    y(\Graph) &= 
    \text{sgn}\left\{\braket{\phi(\Graph)|\phi_0}\right\}\\
    &=\text{sgn}\left\{\Sum_{i \in SV} y_i \alpha_i K(\Graph, \Graph_i)\right\}\label{eq:SVMb},
\end{align}
with 
\begin{equation}
    \label{eq:SVMphi0}
    \phi_0 = \Sum_{i \in SV} y_i \alpha_i\phi(\Graph_i)
\end{equation}
In this case, the training of the kernel amounts to finding the optimal feature vector $\phi_0$. It is worth noting that in many cases, equation \eqref{eq:SVMb} is evaluated directly, without explicitly computing $\phi_0$.

If the dataset is to be split into more than two classes, a popular approach is to combine several binary classification in a one-vs-one scheme\cite{Pawara20}. This means that a classifier is constructed for each pair of classes in the dataset. Namely, for a dataset with $n_c$ classes,  $n_c(n_c - 1)/2$ classifiers will be constructed and trained (one for each pair of classes). This is the strategy that will be used here, whenever necessary.

\subsection{Kernel Ridge Regression}

The regression is similar to the classification task, but here the aim is to attribute a {\it continuous} value to each graph.
Given a training graph dataset $\left\{\Graph_i\right\}_{i=1\ldots M}$, and a set of labels ${\bf y}=\left\{y_i\right\}$ (that we assumed here to be in $\mathbb{R}$), the problem of linear regression consists in finding weights 
$\alphab=\left\{\alpha_i\right\}_{i=1\ldots d}$ (where $d$ is the dimension of the embedding space $\mathcal{H}$),
such that for any new input graph $\Graph$, $y_{\alphab}(\Graph) = \alphab^T\phi(\Graph)$. 
The solution $\alphab$ is found by minimizing
\begin{equation}
    \label{eqn:func_reg}
    J(\alphab) = \frac{1}{2} \Sum_{i=1}^{M} \left[y_{\alphab}(\Graph_i) - y_i\right]^2 + \frac{\lambda}{2}\alphab^T\alphab,
\end{equation}
where $\lambda$ is the regularization hyperparameter.
This problem has a dual formulation by setting ${\bf a} = \boldsymbol{\Phi}^T\alphab$ where $\boldsymbol{\Phi}$ is the matrix whose rows are the embedded vectors $\phi(\Graph_i)$. By injecting the value of ${ \bf a}=\{a_i\}_{i=1\ldots M}$ in \eqref{eqn:func_reg} the solution to the problem is given by
\begin{equation}
    {\bf a} = ({\bf K} + \lambda I)^{-1} {\bf y}
\end{equation}
where ${\bf K}=\left\{K(\Graph_i,\Graph_j)\right\}_{ij}=\left\{\braket{\phi(\Graph_i)|\phi(\Graph_j)}\right\}_{ij}$ is the kernel matrix and ${\bf y}$ is the vector of targets.

The prediction for a new input $\Graph$ is then given by:
\begin{equation}
    y(\Graph) =\Sum_{i=1}^{M} a_i K(\Graph, \Graph_i)
\end{equation}

The kernel presented here can be used for both classification and regression tasks, but only the former will be developed here.

\section{ Quantum Evolution Kernel }
\label{sec:QuantumGraphKernel}
We now present a new Quantum Evolution (QE) Kernel, using the dynamics of an interacting quantum system as a tool to characterize graphs.
The approach we propose consists in associating each graph $\Graph$ with 
a probability distribution $\mathcal{P}_\Graph$, obtained by the measurement of an observable on a 
quantum system whose dynamics is driven by the topology of $\Graph$.
The QE kernel between two graphs $\Graph$ and $\Graph'$ is
then given as a distance between their respective probability distributions $\mathcal{P}_{\Graph}$ and $\mathcal{P}_{\Graph'}$.

\subsection{Time-evolution}

The time-evolution of a quantum state on a graph is a rich source of features for machine learning tasks such as the aforementioned classification and regression.
Let us first detail the construction of the time-evolution. 
\begin{figure}[ht!]
    \centering
    \includegraphics[width=1.\linewidth]{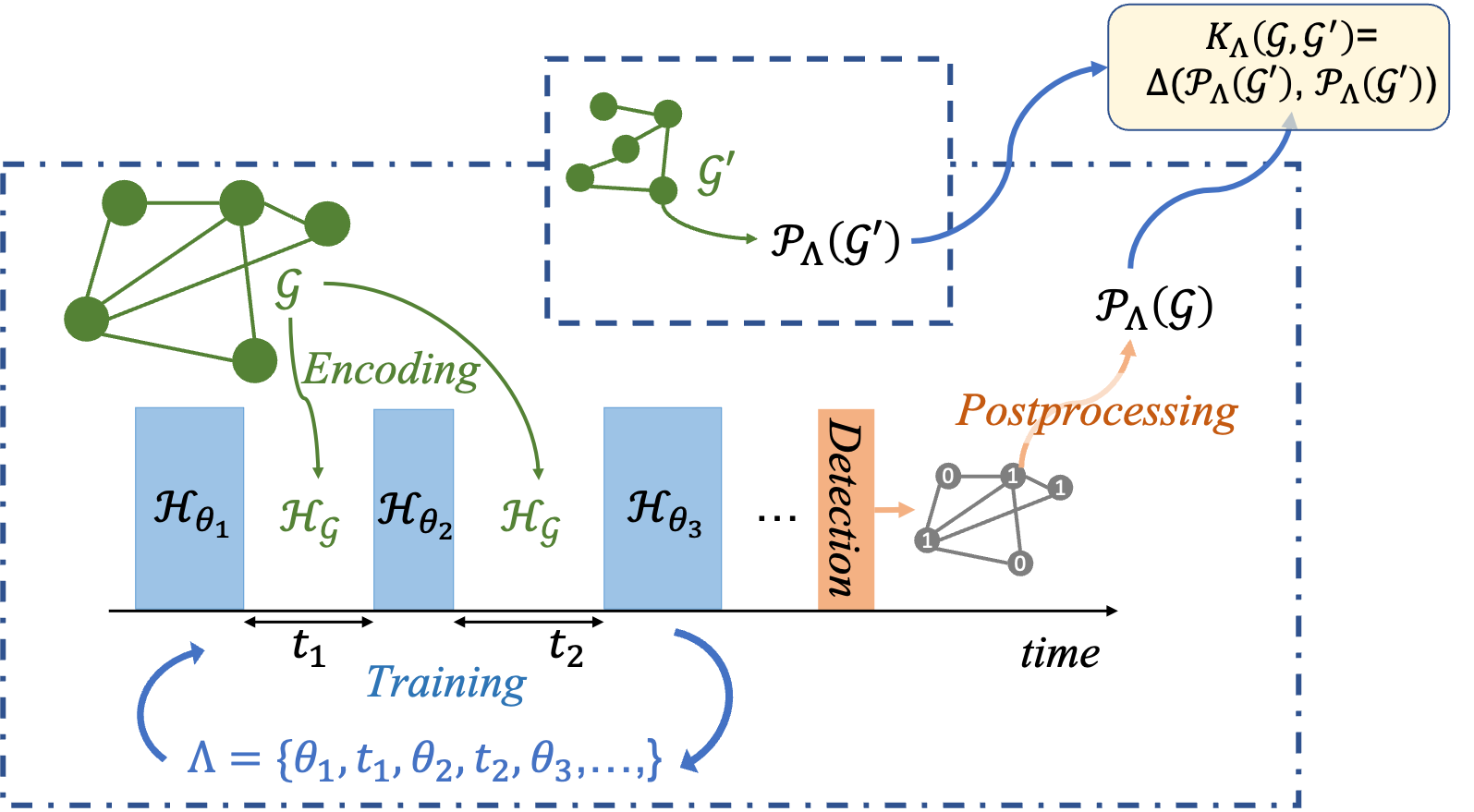}
    \caption{Schematics of the feature map at the heart of  the QE kernel. One first encodes an input graph $\Graph$ into a Hamiltonian $\Ham_\Graph$, 
    which is used in a parametrized sequence, alternating evolution with $\Ham_\Graph$ and pulses with Hamiltonian
    $\Ham_{\theta_i}$. An observable is measured at the end of the pulse, yielding a bitstring. From this bitstring (or a list of bistrings resulting from repeated iterations of this pulse), a probability distribution $\mathcal{P}_\Lambda(\Graph)$ is extracted, out of which the graph kernel is computed, as a distance $\Delta$ on probability distributions. The precise form of the pulse sequence is determined through training on a graph data set.}
    \label{fig:reg_draw}
\end{figure}

We consider a system whose time-evolution is governed by the Hamiltonian $\Ham_\Graph$, 
that can be any parametrized Hamiltonian whose topology of interactions
is that of the graph under study. 
Given a graph $\Graph=(\vertices,\edges)$, one can take for example an Ising Hamiltonian of the form 
$\Ham_\Graph=\Ham_I=\sum_{(i,j)\in \mathcal{E}} \hat\sigma^z_{i}  \hat\sigma^z_{j}$, or the XY Hamiltonian 
$\Ham_\Graph=\Ham_{XY}=\sum_{(i,j)\in \edges} ( \hat\sigma^+_{i}  \hat\sigma^-_{j}+h.c.)$.
Those two Hamiltonians are analyzed here because they are ubiquitous spin models, that can be rather easily implemented on currently existing platforms (particularly in the case of neutral-atom processors, as described in \ref{sec:implementation}). 
Depending on the problem at hand and the features of the graph one is trying to take into account, different Hamiltonians could be used
\footnote{Synthetic Hamiltonians that can't be realized on quantum hardware could even be used in quantum-inspired classical algorithms implementing the QE kernel.}.

We introduce then another Hamiltonian $\Ham_{{\theta}}$, parametrized by a set of parameters $\theta$, independent
of $\Graph$ and such that $\left[\Ham_{{\theta}},\Ham_\Graph\right]\neq 0$, and 
use it to apply {\it pulses} to the system ({\sl i.e.} letting it evolve with a coherent single-qubit driving $\Ham_\theta$
\footnote{In practice, $\Ham_\Graph$ is not completely turned off during the pulses on real hardware implementations, but can be neglected compared to $\Ham_{\theta}$.} 
for a duration $\tau$\footnote{We can consider that $\tau=1$ and include it in the definition of ${{\theta}}$. }).

The system starts in a predefined state $\ket{\psi_0}$. After an initial pulse with $\Ham_{\theta_0}$, we alternatively let the system evolve with 
$\Ham_\Graph$ (for a duration $\tau_i$) and $\Ham_{\theta_i}$ (as illustrated on Fig. \ref{fig:reg_draw}). This time evolution can be summed
up in the set of parameters 
\begin{gather}
\Lambda = \left\{{\theta_0},t_1,{\theta_1},\ldots,t_p,{\theta_p}\right\}.
\end{gather}
In the 
remainder of the text, $p$ will be referred to as the the depth (or the number of layers) of the procedure. After the time-evolution
the system ends up in the state
\footnote{Throughout the paper, we set $\hbar=1$}
\begin{equation}
	\label{eq:layers}
    \ket{\psi_f} = \Prod_{i=1}^p\left(\e^{-\ii \Ham_{\theta_i}}\e^{-\ii \Ham_\Graph t_i}\right)\e^{-\ii \Ham_{\theta_0}}\ket{\psi_0}
\end{equation}

\noindent{\it Remark:} We present here a {\it layered} time-evolution scheme, but the idea can be easily extended to any kind of parametrized time-evolution $\Ham(t)$ such that the final state is
\begin{equation}
    \ket{\psi_f} = \exp\left[-\ii \Int_{0}^{t_f}dt\, \Ham(t)\right]\ket{\psi_0}
\end{equation}
This can be thought of as taking the limit of an infinite number of layers. Analog quantum processing platforms are particularly well suited to the evaluation of this kind of quantities. 
Indded in those systems, some of the parameters of the Hamiltonian can be continuously tuned (and it is often difficult to generate strictly piece-wise constant Hamiltonians).

\subsection{Probability distribution}
Once the system has been prepared in the final state $\ket{\psi_f}$, an observable $\Obs$ is measured, to be used in the construction 
of the feature vector.
We construct here two types of probability distributions that can be easily derived from quantum states, inspired by two standard types of measurements (power spectrum of an operator and histogram of measurements), but other ones could be considered as well.

\subsubsection{Varying pulse sequence}

The approach we consider consists in varying $\Lambda$, measuring $\Obs$
several times  for each $\Lambda$, so as to get $\bar o(\Lambda) = \bra{\psi_f(\Lambda)}\Obs\ket{\psi_f(\Lambda)}$, 
and then construct a probability distribution out of it. 

For example, if all the $\theta_i$ are kept fixed, and $\Lambda$ varies in the duration 
$t = \sum_it_i$, one could define $\mathcal{P}^{\Obs}_\Graph$ as the (normalized) Fourier transform of $\bar o(t)$ :
\begin{gather}
    \mathcal{P}^{\Obs}_\Graph(\Lambda) = (p_1, \dots p_K) \nonumber\\
    p_k = \frac{1}{T}\left|\Int_{0}^{T}dt\, \e^{-2\ii\pi k t/T}  \bar o(t)\right| \label{eq:proba}.
\end{gather}
As will be illustrated later, the Fourier components of $\bar o(t)$ ({\sl i.e.} its power spectrum) can then be used as a fingerprint of $\Graph$.

\subsubsection{Fixed pulse sequence \label{subsubsec:fps}}
Another possible probability distribution is obtained from the tomography of the final state $\ket{\psi_f}$. 
It can be reconstructed from a series of measurements $\{m_1,\ldots,m_M\}$ of the observable $\Obs$ at the end of $M$ repetitions of the exact same pulse sequence.

Let us note $\{\lambda_1, \dots \lambda_K\}$, the eigenvalues of $\Obs$ ({\sl i.e} the possible outcomes
of the measure), and $\{\ket{o_1}, \dots \ket{o_K}\}$ the corresponding eigenstates. 
We can then associate to the graph a probability distribution 
\begin{gather}
    \mathcal{P}^{\Obs}_\Graph(\Lambda) = (p_1, \dots p_K), \text{ where }
    p_k = \left|\braket{o_k|\psi_f}\right|^2.
\end{gather}
By constructing the histogram of the obtained values, one then effectively reconstructs the components of $\ket{\psi_f}$ in the eigenbasis of the operator.

\noindent{\it Remark:} Note that if some eigenvalues are degenerate, one would get instead $p_k = \Sum_i\delta(\lambda_i-\lambda_k)\left|\braket{o_i|\psi_f}\right|^2.$

In practice, if $K$ is large, one would resort to {\it binning} the values of $\lambda_i$, {\sl i.e.} by defining a set of $K'<K$ intervals $\{I_k=[\tilde\lambda_k,\tilde\lambda_{k+1}]\}_{k=1\ldots,K'}$, with $\tilde\lambda_1\leq\lambda_1$ and $\tilde\lambda_{K+1'}\geq\lambda_K$, such that $\mathcal{P}^{\Obs}_\Graph(\Lambda)=(\tilde p_1, \dots \tilde p_{K'})$, where
\begin{gather}
    \tilde p_k  = \frac{|\{ m_i |m_i\in I_k\}|}{M}
                \stackbin[M\to\infty]{}{\equiv} \Sum_{ i |\lambda_i\in I_k}
                \left|\braket{o_i|\psi_f}\right|^2.
\end{gather}

\subsection{Graph kernel}

We can now naturally define a graph kernel by computing the distances between the probability distributions. 
There are many choices of distances between probability distributions. We will here use the Jensen-Shannon divergence\cite{Lin91}, which is commonly used in machine learning.

Given two probability distributions $\mathcal{P}$ and $\mathcal{P}'$, the Jensen-Shannon divergence can be defined as
\begin{gather}
	JS(\mathcal{P}, \mathcal{P}') = H\left(\frac{\mathcal{P}+\mathcal{P}'}{2}\right) -\frac{H(\mathcal{P})+H(\mathcal{P}')}{2},
\end{gather}
where $H(\mathcal{P})=-\Sum_kp_k\log p_k$ is the Shannon entropy of $\mathcal{P}$. $JS(\mathcal{P}, \mathcal{P}')$ takes values in $\left[0,\log 2\right]$. In particular $JS(\mathcal{P}, \mathcal{P})=0$, and $JS(\mathcal{P}, \mathcal{P}')=\log 2$ is maximal if $\mathcal{P}$ and $\mathcal{P}'$
have disjoint supports. 

For two graphs $\Graph$ and $\Graph'$, and their respective probability distributions $\mathcal{P}$ and $\mathcal{P}'$ (computed
as described above), we define the graph kernel as
\begin{gather}
	\mathcal{K}_\mu(\Graph,\Graph') = \exp\left[-\mu\, JS(\mathcal{P}, \mathcal{P}')\right]\in \left[2^{-\mu},1\right].
\end{gather}
The kernel is then positive by construction. Throughout the paper, we will set $\mu=1$, but it might be helpful to adjust this value to improve the results.

The parameter $\Lambda$ is determined through training on a dataset containing graphs whose class is known. 

In the case of SVM, the outcome of training  consists in the optimal value of $\Lambda$ as well as the corresponding support vector coefficients $\alpha_i$ defining the best hyperplane splitting the data set.
The class of a new graph $\Graph$ is then predicted by computing its probability distribution $\mathcal{P}^{\Obs}_\Graph(\Lambda)$ and then computing its kernel values with respect to graphs in the training dataset, according to \eqref{eq:SVMb}.

\section{Ising Quantum Evolution kernel at depth \texorpdfstring{$p=1$}{p=1}\label{sec:Ising}} 

As an illustration, we now derive an expression in a simple case, inspired by Ramsey interferometry, where the probability distribution is the Fourier
transform of the total occupation of an Ising system, after a {\it Ramsey sequence}, comprising a free evolution of duration $t$ between two conjugate global pulses $\Ham_\theta$ and $\Ham_{-\theta}=\Ham_{\theta}^*$.
This example will highlight how relevant characteristics of a graph can be captured, 
even in this very basic setting, by acting uniformly on all the qubits encoding it.

\subsection{Description of the Kernel}
Let us consider a graph $\Graph = (\vertices,\edges)$ and a set of $N=|\vertices|$ qubits interacting with the Ising Hamiltonian
$\Ham_\Graph$, and subject to the the {\it mixing} Hamiltonian $\Ham_\theta$, characterized by a single parameter $\theta=\{\vartheta\}$, defined as
\begin{align}
	\label{eq:Ising}
	\Ham_\Graph=\Sum_{(i,j)\in \mathcal{E}} \hat\sigma^z_{i} \hat\sigma^z_{j} \equiv\Ham_I\text{ and }
	\Ham_\theta = \vartheta \Sum_{i\in\vertices} \hat\sigma^y_i \equiv \vartheta \Ham_1,
\end{align}
where $\hat \sigma^y$ and $\hat \sigma^z$ are the Pauli spin operators. 
The qubit are initially in the {\it empty} product state
\begin{gather}
	\ket{\psi_0} = \bigotimes_{i\in\vertices} \ket{0} \equiv \ket{0\ldots0},
\end{gather}
and we consider the case of a single layer $p=1$, with parameters $\Lambda = \{\theta,t,-\theta\}$,
as defined in \eqref{eq:layers}, so that the final state of the system is
\begin{gather}
	\ket{\psi_\vartheta(t)} \equiv \ket{\psi_f(\Lambda)} = \e^{\ii \vartheta \Ham_1}
						\e^{-\ii t\Ham_\Graph}
						\e^{-\ii \vartheta \Ham_1}\ket{\psi_0}.
\end{gather}
We choose here to measure the total occupation of the system 
$\Obs \equiv \hat n = \sum_{i\in\vertices}(1+\sigma_i^z)/2$, so that
\begin{gather}
	\bar o(t) \equiv n_\vartheta(t) =  \bra{\psi_\vartheta(t)} \hat n \ket{\psi_\vartheta(t)}
\end{gather}
Finally, the probability distribution is computed according to \eqref{eq:proba}
\footnote{$\mathcal{P}^{\Obs}_\Graph(\Lambda)=\mathcal{P}^{\hat n}_\Graph(\theta)$ is defined up to a normalization constant
.}.
In practice, the Fourier transform of the output of the quantum device would be computed over a finite time-interval $T$.
But we will consider here that it is large enough (as compared to the spectrum of the model), and focus on the $T\to\infty$
limit.
In such a procedure, the first pulse brings each qubit in a coherent superposition of the two basis states. During the free evolution of this state, different components of the superposition acquire relative phases, which depend, for each qubit, on the number of neighbors of the corresponding node of the underlying graph $\Graph$. The final pulse maps coherences onto populations, and the analysis of the signal in Fourier space then enables to probe the connectivity structure of the graph, as we show below.

\subsection{Analytical results}

The diagonal nature of the Ising Hamiltonian in the computational basis makes it possible to reach closed formulas, and this should serve as a nice illustrative example.
Details of the calculations can be found in \ref{sec:appendixp1Ising}. 

If a graph $\Graph$ contains $m_\Graph(\kappa)$ vertices of degree $\kappa$, then $\bar o(t)$ reduces to
\footnote{This sum actually finite as $m_\Graph(\kappa)=0$ for any $\kappa > \kappa_{max}$.}
\begin{gather}
	\label{eq:ndensity}
n_\vartheta(t)=2\,{\cos^2\vartheta\sin^2\vartheta}\Sum_{\kappa\geq0} m_\Graph(\kappa) w_\kappa(t),\\
\text{with } w_\kappa(t)={\Re\left\{1-\left(\cos^2\vartheta+\e^{\ii  t}\sin^2\vartheta\right)^{\kappa}\right\}}.
\end{gather}
The probability distribution is then given by
\begin{gather}
	\label{eq:pdensity}
	p_0^{(\infty)}=2\,{\cos^2\vartheta\sin^2\vartheta}\Sum_{\kappa\geq0} m_\Graph(\kappa)\,(1-\cos^{2\kappa}\vartheta),\\
	p_{k>0}^{(\infty)}={(\Sin)^{2(1+k)}}\Sum_{\kappa\geq k} \binom{\kappa}{k}m_\Graph(\kappa)\,(\Cos)^{2(\kappa+1-k)}. \nonumber
\end{gather}
This can be very efficiently computed. Indeed, since $m_\Graph(\kappa)=0$ for $\kappa>\kappa_{\max}$, where $\kappa_{\max}<N-1$ is the maximum degree of the graph, one can introduce a set of $N$ vectors 
$\{{\bf V}_k\}_{k=0,\ldots,N-1}$, of dimension $N$, defined as
\begin{align*}
	{\bf V}_0&=	2\cos^2\vartheta\sin^2\vartheta\left\{1-\cos^{2\kappa}\vartheta\right\}_{\kappa\geq0} \\
	{\bf V}_{k>0}&=	(\Cos)^{2(1-k)}(\Sin)^{2(1+k)}\left\{\binom{\kappa}{k}(\Cos)^{2\kappa}\right\}_{\kappa\geq0}.
\end{align*}
For a given graph $\Graph$, one then just needs to compute the degree histogram vector 
\begin{equation}
	{\bf m}_\Graph = \left\{m_\Graph(\kappa)\right\}_{\kappa\geq0},
\end{equation}
so that the components of the feature vector are obtain via a simple scalar product :
\begin{gather}
	p_{k}^{(\infty)}={\bf m}_\Graph . {\bf V}_{k}.
\end{gather}
\begin{figure*}[t!]
    \centering
    \includegraphics[width=1\linewidth]{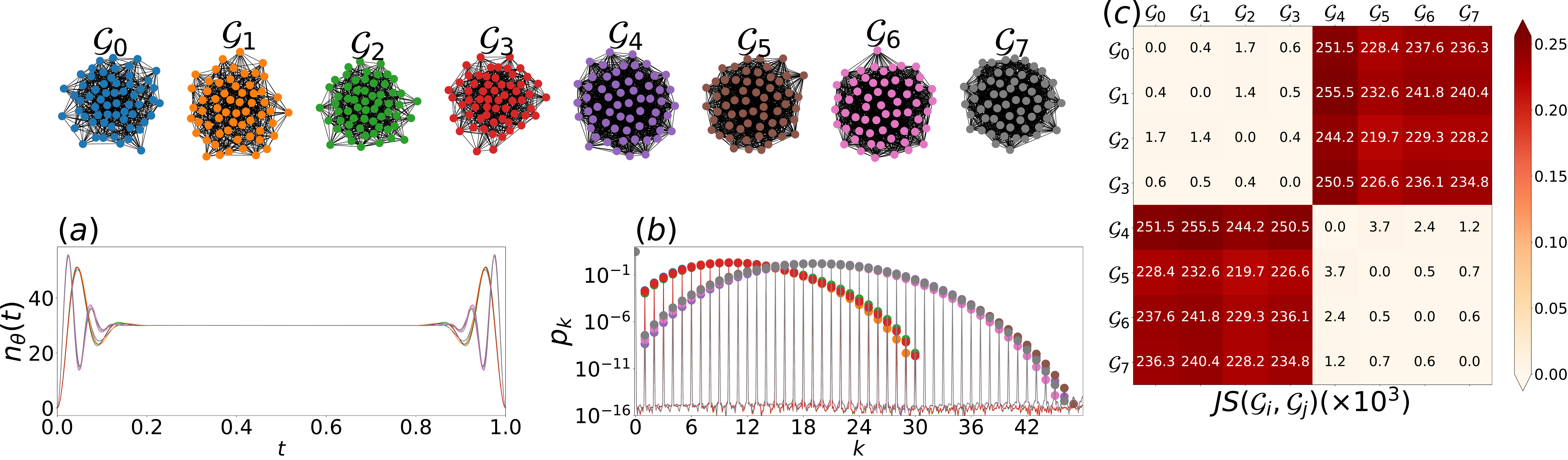}
    \caption{(color online) Illustration of the Ising QE kernel at depth $p=1$, for 8 random Erdős–Rényi graphs of $N=60$ nodes, with edge probability respectively 
    $\rho=.35$ ($\Graph_0,\ldots,\Graph_3$) and $\rho=.65$ ($\Graph_4,\ldots,\Graph_7$), and $\vartheta=\pi/4$. 
    (a) One period of the time dependence of the density as expressed in \eqref{eq:ndensity}, the color of each graph is the same as the corresponding plot.
    (b) Probability distribution extracted from this procedure. The lines are the Fourier Transform of $n_\theta(t)$ (with a finite value of the total time $T$), while the
    dots are given by \eqref{eq:pdensity}. Two types of distributions are clearly emerging, corresponding to the two classes of graphs.
    (c) The Jensen-Shannon divergence computed from $p_k$. The color represents the value of the Jensen-Shannon divergence between the two corresponding graphs (it is 0 on the diagonal, by definition). Bright colors correspond to similar graphs, while darker (red) color correspond to larger values of the divergence (and therefore dissimilar graphs). This groups nicely the graphs of the same class.}
    \label{fig:exple1}
\end{figure*}

The results are illustrated in Fig. \ref{fig:exple1}, for two small sets of graphs generated with the Erdős–Rényi model\cite{Erdos59}, for different edge probability $\rho=0.35$ and $\rho=0.65$. 
The different classes of graphs are nicely captured in this simple case (as one would expect, since the Erdős–Rényi parameter $p$ can be accurately evaluated
from the degree histogram, when the graphs become large enough). Indeed, the distance obtained between graphs of the same category is 100 times smaller than between graphs of different classes.

This simple formula may seem to somehow defeat the purpose of a quantum graph kernel (as no speedup is to be gained from a quantum machine). But this is only a very simple example. In a more generic case --- and in particular in the case described in \ref{sec:Benchmarking} --- the evolution would be more intricate and no simple analytic formula is expected to be computable.

\section{Connection to random-walk kernels \label{sec:ExistingKernels}}
	
Before we delve into the benchmarking of the QE kernel on a specific case in \ref{sec:Benchmarking}, 
let us illustrate how it is very general, specifically how it can be connected to already existing kernels, with a proper choice of the Hamiltonian and/or observable.
	
\subsection{Steady-state random walk kernel}
Random-walk based kernels can be more efficiently computed than graphlet-based ones\cite{Shervashidze11}, and have therefore
been extensively used in the literature. We show here how the QE kernel can be linked to these kernels.

We still work in the case of the Ising Hamiltonian, but we now measure separately the occupation of each site ({\sl i.e.} $\Obs=\hat{n}_i$), and computes the probability distribution $\mathcal{P}_\Graph(\theta)=\left\{p^{(i)}_0\right\}_{i\in\vertices}$, with $p_0^{(i)}\equiv\overline{n_i(t)}$ the time-averaged occupation of site $i$. It can be expressed as
\begin{gather}
	p_0^{(i)}=2\,{\cos^2\vartheta\sin^2\vartheta}\,(1-\cos^{2\kappa_i}\vartheta)
	\underset{\vartheta\to0}{\approx}2\vartheta^4\left\{\kappa_i\right\},
\end{gather}
where $\kappa_i$ is the degree of vertex $i$.

This distribution is then equivalent to that of the classical steady-state random walk\cite{Bai_thesis}, in which the time spent on each vertex (or the probability of visiting it) is used as a component of the feature vector.

\subsection{Quantum walks}
    It has be shown that, for certain types of graphs, a classical random-walk fails to explore efficiently the system,
    while a quantum walk would be able to do it \cite{Childs03}.
	We now briefly examine the case where the graph Hamiltonian is chosen to be $H_{XY}$, and highlight its connection
	to random-walk kernels.

\subsubsection{Quantum-walk kernel}

The XY Hamiltonian $\Ham_{XY}$ preserves the total occupation number. We introduce 
$H_n = \text{span}\left\{\ket{\sigma} | \sum_{i\in\vertices}\sigma_i=n\right\}$ the subspace 
of states with occupation $n$. $\Ham_{XY}$ is then block-diagonal, with each block acting on one of the $H_n$. We will note
$\Ham_n$ the restriction of $\Ham_{XY}$ to this subspace.

In particular, the matrix representing $\Ham_1$ is (isomorphic to) the adjacency matrix of $\Graph$. If one is able to prepare the system in an initial state
$\ket{\psi_0}$ that only has components in $H_1$, the evolution of the system with $\Ham_{XY}$ then corresponds to a random walk of a single (potentially delocalized) particle on the graph. If one adds a longitudinal field term $\sum_i \kappa_i \hat n_i$ (with $\kappa_i$ the degree of vertex $i$.), the Hamiltonian becomes the Laplacian of the graph. 
From the dynamics of this system, one can then reconstruct the {\it quantum-walk kernel}\cite{Rossi15,Bai15}, from the histograms of the local density operators $\hat n_i$.

\subsubsection{Generic occupation graphs}
The $XY$ Hamiltonian presents further applications. Let us consider again a graph $\Graph=(\vertices,\edges)$, containing $N=|\vertices|$ nodes and $M=|\edges|$ edges.
For any value of $n$,  $\Ham_n$ describes the random walk on $\Graph$ of $n$ particles with hardcore interactions. 
Let us define an effective graph $\Graph_n$, with $\binom{N}{n}$ vertices corresponding to the $\binom{N}{n}$  Ising configurations $\sigma$ with $n$ 1s, such that $(\sigma,\sigma')\in \edges_n$ if $\braket{\sigma |\Ham_n|\sigma'} = 1$. The matrix representing $\Ham_n$ is then (isomorphic to) the adjacency matrix of $\Graph_n$.

\begin{figure}[tpb]
	\centering
	\includegraphics[width=\linewidth]{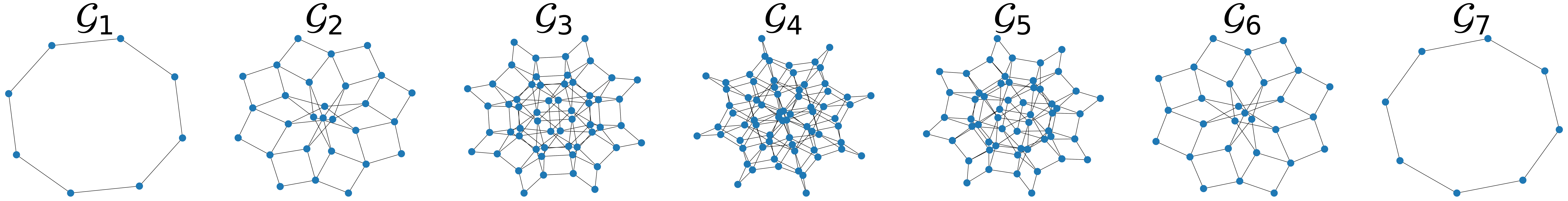}
	\includegraphics[width=\linewidth]{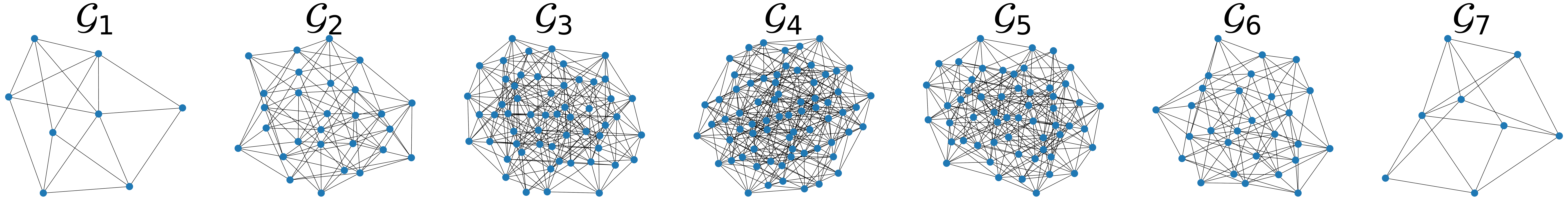}
	\caption{The sequence of occupation graphs $\Graph_n$ from a loop ({\it top row}) and random ({\it bottom row}) initial graph $\Graph\equiv\Graph_1$. As the occupation is increased, both the number of vertices and the number of edges grow exponentially. In the case of the periodic chain, the $\Graph_n$s retain the rotational/permutation symmetry of $\Graph$.}
	\label{fig:occupationgraphs}
\end{figure}

We show the various $\Graph_n$ for a periodic chain graph as well as for an arbitrary graph, both of size $N=8$ (Fig. \ref{fig:occupationgraphs}). 
The computation of the Quantum Evolution kernel is very difficult for a classical computer, as the size of $\Graph_n=(\vertices_n,\edges_n)$ grows quickly with $n\leq N/2$\footnote{Because of the particle-hole symmetry
$\Graph_n$ and $\Graph_{N-n}$ are equivalent.} :
\begin{gather}
|\vertices_n|=\binom{N}{n}\text{ and } 
|\edges_n|=M\binom{N-1}{n-1}=\frac{M n}{N}|\vertices_n|.
\end{gather}
 
 On a quantum device, however, if one is able to prepare the system in a state with a defined occupation $n$, it is then straightforward to compute the QE kernel for one of the $\Graph_n$. One would just need to make sure to use a mixing Hamiltonian $\Ham_\theta$ that also preserves the occupation number ({\sl e.g.} by combining $\Ham_{XY}$ with a longitudinal-field term).
 
 It is worth noting that for a given graph $\Graph$, the density of $\Graph_n$ --- defined as $d_n=2|\edges_n|/|\vertices_n|(|\vertices_n|-1)$ --- decays rapidly as $n\to N/2$.

\subsubsection{Full Hamiltonian dynamics}
	In the case where the initial state overlaps with several occupation number sectors, it is harder to get a proper description
	in terms of random walks, highlighting the difference between classical and quantum walks. 
	However, a link to the random-walk properties of the different $\Graph_n$ can still be made.
	We will illustrate this in a setting similar to that of \ref{sec:Ising}, with the depth set to $p=1$,
	and the parameters $\Lambda = \{\theta,t,-\theta\}$, and the graph Hamiltonian $\Ham_\Graph = \Ham_{XY}$. 
	The observable that is	measured is again the total occupation $\Obs = \sum_i\hat n_i$. After a sequence with
	a free evolution of duration $t$, the expected total occupation is given by
\begin{multline}
	\label{eq:noftXYseries}
				n(t) = 2N(\Cos)^{2}(\Sin)^2\\
				- 2(\Cos)^{2N+2}\Sum_{\sigma\in\{0,1\}^N}\varrho_{n_\sigma}(t),
\end{multline}
with
    \begin{multline}
		\varrho_{n_\sigma}(t)
		=(\Tan)^{2n_\sigma}\Sum_{\substack{p,q\geq0\\\delta=0,1}}\left\{
		{(-1)^{\delta}(-t^2)^{p+q+\delta}}\right.\\
		\left.\times \frac{l_{2p+\delta}(\sigma)}{(2p+\delta)!}\Sum_{\substack{\tilde\sigma \in \mathcal{F}^-_{\sigma}}}
		\frac{l_{2q+\delta}(\tilde\sigma)}{(2q+\delta)!}\right\},
    \end{multline}
    where $l_{p}(\sigma)$ is the number of random walks of length $p$ in $\Graph_{n_\sigma}$ starting at $\sigma$, 
    and $\mathcal{F}^-_\sigma =\left\{\tilde\sigma\in\Graph_{n_\sigma-1}|\braket{\tilde\sigma|\hat\sigma^-_i|\sigma}=1\right\} $ is the set
    of configurations obtained by flipping one of the $\sigma_i=1$ to 0.
    This expression combines properties of random-walks on different occupation graphs, illustrating how various occupation sectors are involved in QE kernels based on the $XY$ Hamiltonian.
    
    One may then expect the resulting classifier to capture intricate properties that a random-walk on a single sector or a classical walk would miss.
    
\section{Benchmarking of the Quantum Evolution Kernel\label{sec:Benchmarking}}

The scheme presented in \ref{sec:Ising} requires the measurement of the time-dependence of the expectation value of an operator.
In most existing quantum computing platform, this would require a prohibitive number of measurements, in order to have sufficient statistics on enough Fourier components.
In order to circumvent this issue, we now turn to a more promising scheme, requiring fewer measurements. In this setting, on reconstructs the histogram of the observable in the final state, in a {\it fixed pulse
sequence}, as briefly mentioned in \ref{subsubsec:fps}.

We consider again a $\Graph = (\vertices,\edges)$ and a set of $N=|\vertices|$ qubits. 
The observable to be measured is the {\it Ising energy} associated with the graph :
\begin{gather}
	\Obs = \Ham_I=\Sum_{(i,j)\in \mathcal{E}} \hat\sigma^z_{i} \hat\sigma^z_{j}
\end{gather}

The goal is to sample $\Obs$ in the final state $\ket{\psi_f}$, obtained after $p$ layers :
\begin{equation}
    \ket{\psi_f(\Lambda)} = \Prod_{i=1}^p\left(\e^{-\ii \Ham_{\theta_i}}\e^{-\ii \Ham_\Graph t_i}\right)\e^{-\ii \Ham_{\theta_0}}\ket{\psi_0},
\end{equation}
with $\Lambda = \left\{{{\theta_0}=\pi/4},t_1,{\theta_1},\ldots,t_p,{\theta_p}\right\}$ fixed for a given dataset, and determined by the training procedure.

By repeating the measurement $\bra{\psi_f}\Obs\ket{\psi_f}$ many times, one reconstructs the tomography of $\psi_f$ in the eigenbasis $\left\{\ket{o_i}\right\}_i$ of $\Obs$. Here, because of the choice of observable, this basis coincides with the computational basis ($\left\{\ket{o_i}\right\}_i=\left\{\ket{\sigma_i}|{\sigma_i\in\{0,1\}^N}\right\}$).
Indeed, for 
\begin{equation}
	\ket{\psi_f(\Lambda)}=\Sum_{i} \psi_f^{(i)}\ket{o_i},
\end{equation}
the result of the measurement of $\Obs$ gives $E_\sigma=\sum_{(i,j)\in \mathcal{E}} \sigma_{i} \sigma_{j}$ with probability $p_i=|\psi_f^{(i)}|^2$.
The probability distribution that this methods tries to reconstruct is closely related to $\mathcal{P}_\Graph(\Lambda) = \left\{\left|\braket{\psi_f(\Lambda)|\sigma_i}\right|^2\right\}_{i}$
\footnote{Note that this is only true if the eigenvalues of $\Obs$ are non-degenerate.}.
It would take an exponentially growing number of measurements to evaluate this exact distribution. However, as the results of this section and of \ref{sec:implementation} will show, sufficient information can be obtained from the histogram of the measured values, when the size of the bins is small enough. For a given bin width $\delta \epsilon$, the resulting probability distribution from a series of $M$ measurements $\{\tilde o_1,\ldots,\tilde o_M\}$ is then
\begin{align}
    \{\mathcal{P}^{(\delta \epsilon)}_\Graph(\Lambda)\}_i &= 
    \frac{1}{M}\Sum_{j} \chi_i(\tilde o_j)\nonumber\\
	&\equiv \Sum_{j} \chi_i(o_j) \left|\braket{\psi_f(\Lambda)|\sigma_j}\right|^2
\end{align}
where $\chi_i$ is the in indicator function of the $i-th$ bin of the histogram.
We now proceed to characterize this kernel. To this end, 
we consider the graph Hamiltonian to be either the Ising Hamiltonian $\Ham_I$ (diagonal in the computational basis),
or the XY Hamiltonian $\Ham_{XY}$ (non-diagonal in the computational basis).
We follow a procedure similar to that described in \cite{Schuld20} to benchmark the graph kernel.

We tested the protocol on a graph classification task on several datasets and we compare the results with other graph kernels. Each graph kernel is associated with a Support Vector Machine, and the score is the accuracy,
namely the proportion of graph in the test set that are attributed the correct class.
    We perform a grid search on the hyperparameter $C$ (defined in \ref{sec:SVM}) in the range $[10^{-3}, 10^{3}]$ and we select the value with the best score.
For each value of the parameters $\Lambda$, the obtained accuracy is averaged over 10 repetitions of the following cross-validation procedure : 
the dataset is first randomly split into 10 subsets of equal size. The kernel is then tested ({\sl i.e.} its accuracy is measured) on each of these 10 subsets, after having been trained on the 9 other subsets. The accuracy resulting from this split is then the average accuracy over the 10 possible choices of test subset.

The datasets are taken from the repository of the Technical University of Dortmund \cite{Morris+2020}. Attributes and labels of nodes and edges are ignored. Because the simulation of large quantum systems is algorithmically costly, only graphs between 1 and 16 nodes are considered. 
In the case of the Fingerprint dataset, only the classes 0, 4 and 5 are considered, because each other class constitute at most $\sim 7 \%$ of the dataset (to be compared to the $\sim 25\%$ of each of the classes 0, 4 and 5). Table \ref{tab:stats} summarizes the characteristics of each dataset after preprocessing.

\begin{table}[ht]
  \centering
  \begin{tabular}{|l||c|c|c|}\hline
    Dataset    			&  samples		&  classes	& samples per classes	\\\hline
    IMDB-MULTI 		& 1185	& 3	& 371, 403, 411\\
    IMDB-BIN 		& 499	& 2 & 239, 260\\
    PTC\_FM 		& 234	& 2	& 135, 99\\
    PROTEINS 		& 307	& 2	& 82, 225\\
    NCI1 			& 361	& 2	& 282, 79\\
    Fingerprint		& 1467	&  3 & 515, 455, 597\\\hline
  \end{tabular}
  \caption{Characteristics of the datasets after preprocessing, keeping graphs containing less than 16 nodes, and classes with enough representatives.}
  \label{tab:stats}
\end{table}

The parameter $\Lambda$ is trained by Bayesian Optimization. The function to optimize is the accuracy score averaged on all cross-validation splits, and a gaussian surrogate function is trained with a Matérn Kernel (see Appendix \ref{app:bayes} for more details). At each step, the surrogate function is minimized by sampling 5000 points. The computation time depends on the dataset and the type of Hamiltonian used. The number of evaluations is then chosen so as to keep the total evaluation times of the order of a few hours across datasets.

We compared the QE kernels to the Graphlet Subsampling (GS) kernel\cite{Przulj07,pmlr-v5-shervashidze09a} and the Random Walk (RW) kernel\cite{Vishwanathan10}. The procedure used to compute these kernels are described in \ref{sec:appendix classical kernels}. In the GS kernel, we used 1000 samples and perform a grid search over the graphlet subsizes from 3---6. For the RW kernel, we performed a grid-search over the weight hyperparameter in $[0.001, 0.01]$.  The values of the penalty hyperparameter $C$ have been restricted to $[10^{-3}, 10^{-1}]$ to limit the computation time. The simulations of classical kernels have been done with the \texttt{grakel} library \cite{JMLR:v21:18-370}.

\begin{table*}[ht]
  \centering
  \begin{tabular}{|l||c|c|c|c|c|c|}\hline
    Dataset    			& Ising$_1$	(150)	& Ising$_4$ (2000)		& Ising$_8$	(6000)		& XY$_4$ (2000)			& GS	& RW     \\\hline
    IMDB-MULTI 		& $46.8\pm4.4$		& ${\bf48.1\pm4.4}$ 	& $47.7\pm4.4$		& $47.5\pm4.5$	& $40.9\pm3.5$ 		& $45.2\pm3.4$\\
    IMDB-BIN 			& $69.0\pm6.1$		& $71.6\pm5.7$ 		& ${\bf71.8\pm5.4}$	& $70.6\pm5.6$	& $66.5\pm5.9$ 	& $67.8\pm6.5$\\
    PTC\_FM 			& $62.5\pm7.9$		&$ 65.8\pm7.9$ 		& ${\bf66.0\pm7.6}$	& $65.2\pm8.2$	& $61.5\pm8.9$	& $59.4\pm7.8$ \\
    PROTEINS 			& $73.3\pm 1.2$		& $74.5\pm2.6$	& ${\bf76.0\pm5.3}$		& $74.8\pm3.7$	& $73.3\pm1.2$	& $73.3\pm 1.2$ \\
    NCI1 				& $78.1\pm 0.8$		& $78.6\pm 3.2$ 		& ${\bf80.1\pm3.5}$	& $78.8\pm4.8$	& $78.1\pm0.8$ 	& $78.1\pm0.8$\\
    Fingerprint 			& $58.6\pm2.0$		& ${\bf60.2\pm3.2}$	& $60.1\pm3.3$		& $60.1\pm3.3$	& $57.9\pm3.3$ 	&$59.9\pm2.2$\\\hline
  \end{tabular}
  \caption{Comparison of the accuracy of the graph kernel on various data sets. The average over all cross-validation splits and the standard deviation associated is displayed.
  These values were obtained measuring the Ising energy after $p$ layers
  with an Ising graph Hamiltonian (Ising$_p$) or an XY Hamiltonian (XY$_p$). 
  Next to each model is indicated a rough approximation of the number of evaluations. 
  The quantum kernel in the Ising setting outperforms the other kernels considered.
  The resources needed to evaluate the score on bigger data sets ({sl e.g.}IMDB-MULTI and Fingerprint) allow for less evaluations. Models with more layers could be tested in the case of Ising, which is faster to simulate than XY.
  }
  \label{tab:summary}
\end{table*}

All the results are summarized in table \ref{tab:summary}. For all datasets, at least one quantum kernel is better than the best classical kernel. The depth of the evolution plays an important role, since increasing the number of layers for the Ising evolution almost always leads to a better score. This confirms the fact that adding more layers allows for the capture of richer features. For IMDB-MULTI and Fingerprints, the 8-layer Ising scheme performances seem worse than those of the 4-layer Ising. It is an artifact due to the incomplete convergence of the optimization. Indeed, since those datasets are bigger, less evaluations of the function were allowed in our computational budget.

Our results show that a graph kernel based on the time-evolution of a quantum system can lead to an efficient classification of graphs. Since many different QE kernels can be built, one immediate avenue to improvement would be to try to build a better kernel by combining them. 
This is the idea behind multikernel learning\cite{gonen2011multiple}. The main appeal to this approach here is that it could efficiently combine quantum and classical computing, by determining classically the optimal combination among kernels that were trained on a QPU
(see \ref{app:Multikernel} for more details).
\section{Implementation on neutral-atom platforms\label{sec:implementation}}

Several platforms have been developed recently, aiming at simulating quantum spin models: ions \cite{Monroe21}, polaritons \cite{berloff2017} and neutral atoms \cite{browaeys2020many}. Those quantum processors (QPU) would be particularly well suited to an implementation of the Quantum Evolution kernel.
We detail here the case of a programmable array of qubits made of neutral atoms.

\subsection{Description of the platform}
\begin{figure}
    \centering
    \includegraphics[width=\linewidth]{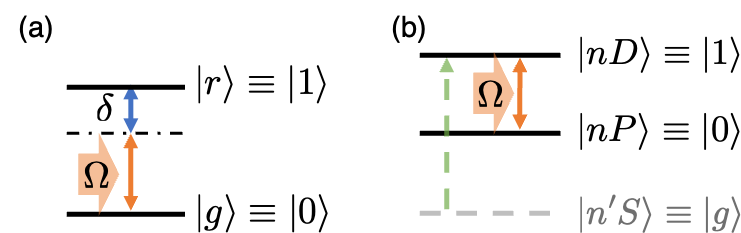}
    \caption{Implementation of quantum spin models with neutral atoms. In the Ising case (a), the spin states $\ket{0}$ and $\ket{1}$ correspond to a ground state $\ket{g}$ and a rydberg state $\ket{r}$ of the atom. The amplitude $\Omega$ of the pumping laser determines the transverse-field term in the Ising model, and the detuning to resonance $\delta$ induces a longitudinal-field term.
    In the XY case (b), the spin states correspond to two dipolar rydberg state $\ket{nP}$ and $\ket{nD}$. The atoms are first brought from a ground state $\ket{g}$ to $\ket{nD}$. 
    A microwave driving between  $\ket{nP}$ and $\ket{nD}$ can additionally generate tunable $\sigma^x$ and $\sigma^z$ terms.
    }
    \label{fig:rydberg}
\end{figure}

Arrays of individual neutral atoms trapped in optical tweezers
\,\cite{Saffman10,Saffman2016,barredo_atom-by-atom_2016,endres_atom-by-atom_2016,barredo_synthetic_2018,browaeys2020many,Henriet2020quantum,Morgado20,Beterov20,Wu21} 
have emerged as a promising platform for quantum information processing. By promoting the atoms to so-called 
Rydberg states, they behave as huge electric dipoles and experience dipole-dipole interactions that map into spin Hamiltonians. 
The spins undergo various types of interactions depending on the Rydberg states involved in the process, leading to different Hamiltonians~\cite{browaeys2020many}, as illustrated in Fig \ref{fig:rydberg}. 
The most common configuration is the Ising model, illustrated in \ref{fig:rydberg}(a) which is obtained when the $\ket{0}$ 
state is one of the ground states $\ket{g}$ and the $\ket{1}$ state a Rydberg state $\ket{r}$\cite{schauss2015crystallization,labuhn2016tunable,Bernien17,leseleuc2018accurate}. 
In this case, a van der Waals interaction between atoms arises due to non-resonant dipolar interactions, and leads to an interaction term of the form 
$C_6/(r_{ij}^6) n_i n_j$, where $n_j=(1+\sigma_j^z)/2$ represents the Rydberg state occupancy of the spin $j$. The coupling between two spins $i$ and $j$ depends 
on the inverse of the distance between them $r_{ij}$ to the power of 6, and on a coefficient $C_6$ relative to the Rydberg state.

The XY Hamiltonian is another example of a spin model that one can realize on this platform. It naturally appears when the spin states $\ket{0}$ 
and $\ket{1}$ are two Rydberg states with a finite dipolar moment (such as {\sl e.g.} $\ket{nP}$ and $\ket{nD}$~\cite{barredo2015coherent,orioli2018relaxation}, or $\ket{nS}$ and $\ket{nP}$~\cite{deleseleuc19}). In that case, 
a resonant dipole-dipole interaction occurs, resulting in the creation of a term $2C_3/(r_{ij}^3) \left(\sigma_i^x \sigma_j^x + \sigma_i^y \sigma_j^y \right)$ between atoms. 
This term describes a coherent exchange of spin states, transforming the pair state $\ket{01}$ into $\ket{10}$. 
This interaction scales as the inverse of the distance $r_{ij}$ to the power of 3 and depends on a coefficient $C_3$ relative to the Rydberg states.

A driving field can be used to implement the rotations. 
By tuning the amplitude, the phase and the frequency of the driving, one can engineer any {\sl global} single-qubit operation (or, in terms of spin models, any uniform magnetic field ${\boldsymbol\Omega}$ so as to add a term
$\sum_{i\in\vertices}\boldsymbol\Omega.(\hat\sigma_i^x,\hat\sigma_i^y,\hat\sigma_i^z)$ to the Hamiltonian).

As the interaction strength depends on the distance between atoms, only local graphs 
can be embedded in this platform. As such, only databases containing local graphs could be used. Note that an additional position-dependent detuning 
term could be used for graphs with labeled nodes.

\subsection{Illustration on a simple dataset}
As an example, we emulate the computation of the graph kernel described in \ref{sec:Benchmarking}, and apply it to a simple dataset containing graphs that are small enough to be computationally tractable with classical computers, ahead of implementation on a real hardware.

These simulations are done using the  \texttt{pulser} library\cite{Pulser}. In these, the characteristics of the laser acting on the atoms (amplitude and frequency) are set at each time, tuning the Hamiltonian seen by the atoms. 
The interaction Hamiltonian is determined by the spatial distribution of the atoms, that can be tuned to any arbitrary geometry (current hardware implementation is limited to 2D, but even 3D configurations could be assembled similarly).

\begin{figure*}[htb!]
\includegraphics[width=\linewidth]{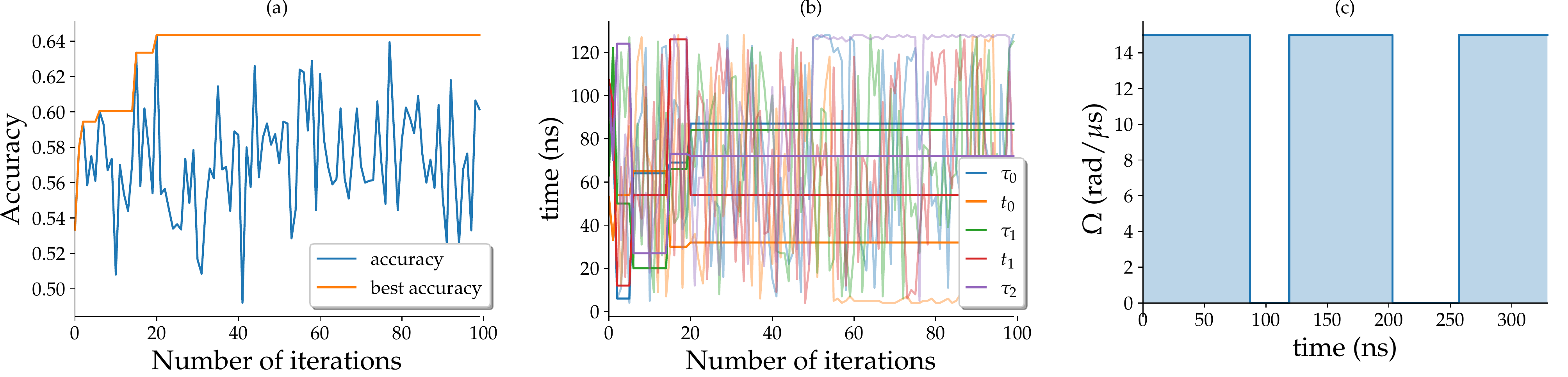}
\caption{(a) Accuracy of the quantum graph kernel, in the case of the Ising Graph Hamiltonian, with $p=2$ layers, as a function of the number of bayesian optimization, with a false positive and false negative detection error of $\epsilon=0.05$ and $\epsilon^\prime=0.05$ respectively. The blue curve is the accuracy obtained from the parameters set a each iteration, while the orange curve represents the best accuracy obtained up to each iteration. 
After $\sim20$ iterations, the accuracy of the kernel goes from $\sim.55$ to $\sim.64$, its maximal value reached in this case. Further iterations don't lead to any improvement.
(b) The corresponding values of the parameters $\Lambda$ of the pulse. Faint lines are the values at each iteration while the solid lines are the values reaching the best accuracy. The values keep fluctuating, suggesting that it may be possible to further train the kernel.
(c) Optimal amplitude $\Omega$ of the pulse sequence obtained after training on Fingerprint*.}
\label{fig:pulser}
\end{figure*}

\subsubsection{Details of the implementation}

We focus here on the dataset Fingerprint. 
We used for that purpose a subset (referred to as Fingerprint* in the remainder of the text) consisting of 200 graphs from this dataset, restricted to graphs up to 12 nodes. As before, only the categories 0, 4 and 5 were considered.

With the notation presented in \ref{sec:QuantumGraphKernel}, the chosen graph and mixing Hamiltonians are
\begin{gather}
    \Ham_\Graph = \Sum_{i>j}J_{ij}\hat\sigma_i^z\hat\sigma_j^z\hspace{.5cm}
    \Ham_\theta = \tau \Ham_\Graph + \frac{\Omega_0\tau}{2}\Sum_{i\in\vertices}\hat\sigma_i^x,
\end{gather}
where here the coupling $J_{ij} = C_6/R_{ij}^6$ coupling between atoms $i$ and $j$ decays with the distance $R_{ij}$ between them. The dataset contains coordinates of the nodes such that the graphs are locals, we used them to position the atoms in the register. A rescaling is applied such that the minimum distance between the atoms is 5 $\mu$m. 
The amplitude is kept constant and set to the highest value reachable by the QPU ($\Omega_0=15$ rad/$\mu$s). The detuning is set here to $\delta=0$, so that a mixing pulse is then characterized by a single parameter ${\theta}_i = \tau_i$.
 We chose an architecture of 2 layers of alternating Rabi pulses ($\Omega=\Omega_0$, $\delta=0$) and free evolution, so that $\Lambda = \left\{\tau_0,t_0,\tau_1,t_1,\tau_2\right\}$.
The values of the durations $\left\{\tau_0,t_0,\tau_1,t_1,\tau_2\right\}$ are the variables to be optimized. In order to get a realistic pulse for the existing hardware, with a finite response and coherence times, we impose that  $\tau_0,t_0,\tau_1,t_1,\tau_2>4$ ns and  $T=\tau_0+t_0+\tau_1+t_1+\tau_2< 500$ ns.

We consider first the case where the various sources of noise are ignored,
and determine the optimal values for $\Lambda$. 
Using the same cross-validation scheme as in \ref{sec:Benchmarking}, we determine the obtained accuracy of the kernel, and compare it to the GS and RW kernels on the same dataset, as summarized in table \ref{tab:pulser}. 
The standard deviation on the accuracy is estimated via cross-validation, similarly to what was done in \ref{sec:Benchmarking}. As before, the quantum kernel reaches a higher accuracy than the reference kernels.

\subsubsection{Sensitivity to noise}

One of the key problematic of quantum computing is the robustness to noise. 
In order to quantify the effect of the noise, we set the false positive and false negative error rates to $\epsilon=0.05$ and $\epsilon^{\prime}=0.05$ respectively (in accordance to conservative estimates for currently existing devices). This type of noise is independent of the duration of the evolution, and for short enough times (less than a few microseconds), it is expected to dominate over other sources of imperfection (decoherence in particular).

In a real implementation, the probability distribution would be reconstructed from a finite number of noisy measurements.
In order to emulate this effect, we estimate the kernel 100 times on different sets of 10000 measurements. For each of these 100 estimations, the accuracy is again evaluated via the same 10-fold cross-validation procedure as before.
The average accuracy is reported in \ref{tab:pulser}, alongside those of other kernels on the same dataset.
With this kind of noise, the estimated accuracy is not significantly altered as compared to the noiseless case, and still outperforms the two comparison kernels.

\begin{table}[ht]
  \centering
  \begin{tabular}{|l||c|c|c|c|}\hline
    Dataset    			& QE(0) & QE(0.05)			& 	 GS			& RW     \\\hline
    Fingerprint* 			& ${\bf64.0\pm8.7}$&  $63.0$  	& $48.2\pm8.9$ 	& $52.0\pm5.8$ \\\hline
  \end{tabular}
  \caption{Accuracy reached on the reduced data set Fingerprint*, by the quantum evolution kernel in the case of a 2-layer scheme (as described in the text) without noise (QE(0)) and with detection error (QE(0.05)), as well as that of the Graphlet Subsampling (GS) and  Random Walk (RW) kernels.
  The average of accuracy over the splits and the associated standard deviation are noted (see table \ref{tab:summary}). For the noisy kernel, the value is the average over all 100 kernel estimates, and thus the standard deviation cannot be directly compared to that of other the kernels in the table.
  }
  \label{tab:pulser}
\end{table}

Let us now briefly analyze the convergence of the optimization.
As illustrated in figure \ref{fig:pulser} (a), most of the accuracy increase is obtained after only a few iterations of optimization ($\sim20$ in the example illustrated here). It might be possible to reach a higher value, but at the cost of much longer computations. This would go beyond the scope of such an illustrative study, and the value reached here already surpasses the GS and RW kernels. 
The corresponding parameters of $\Lambda$ can be seen in figure \ref{fig:pulser}(b) The resulting best pulse is shown in figure \ref{fig:pulser} (c), consisting of three mixing pulses of similar duration (87, 84 and 72 ns respectively, corresponding to pulses with $\vartheta\approx 0.4\pi$, $0.4\pi$ and $\pi/3$ in the non-interacting case described in \ref{sec:Benchmarking}), alternating with shorter free evolutions (32 and 54 ns).

Let us now turn to the alteration of the kernel due to detection errors.
For each pair of graphs $\Graph_i$ and $\Graph_j$, the resulting value of the kernel $K_{\epsilon,\epsilon^\prime}(\Graph,\Graph^\prime)$ averaged over several measurements is
compared to what was obtained in the noiseless case.
To this end, the relative difference between those values is introduced :
    \begin{gather}
        \label{eq:Kij}
        \delta K_{ij}(\epsilon,\epsilon^\prime) = 
        \left|1-\frac{K_{\epsilon,\epsilon^\prime}(\Graph,\Graph^\prime)}
        {K_{0,0}(\Graph_i,\Graph_j)}\right|.
    \end{gather}

\begin{figure}[htpb]
	\centering
	\includegraphics[width=\linewidth]{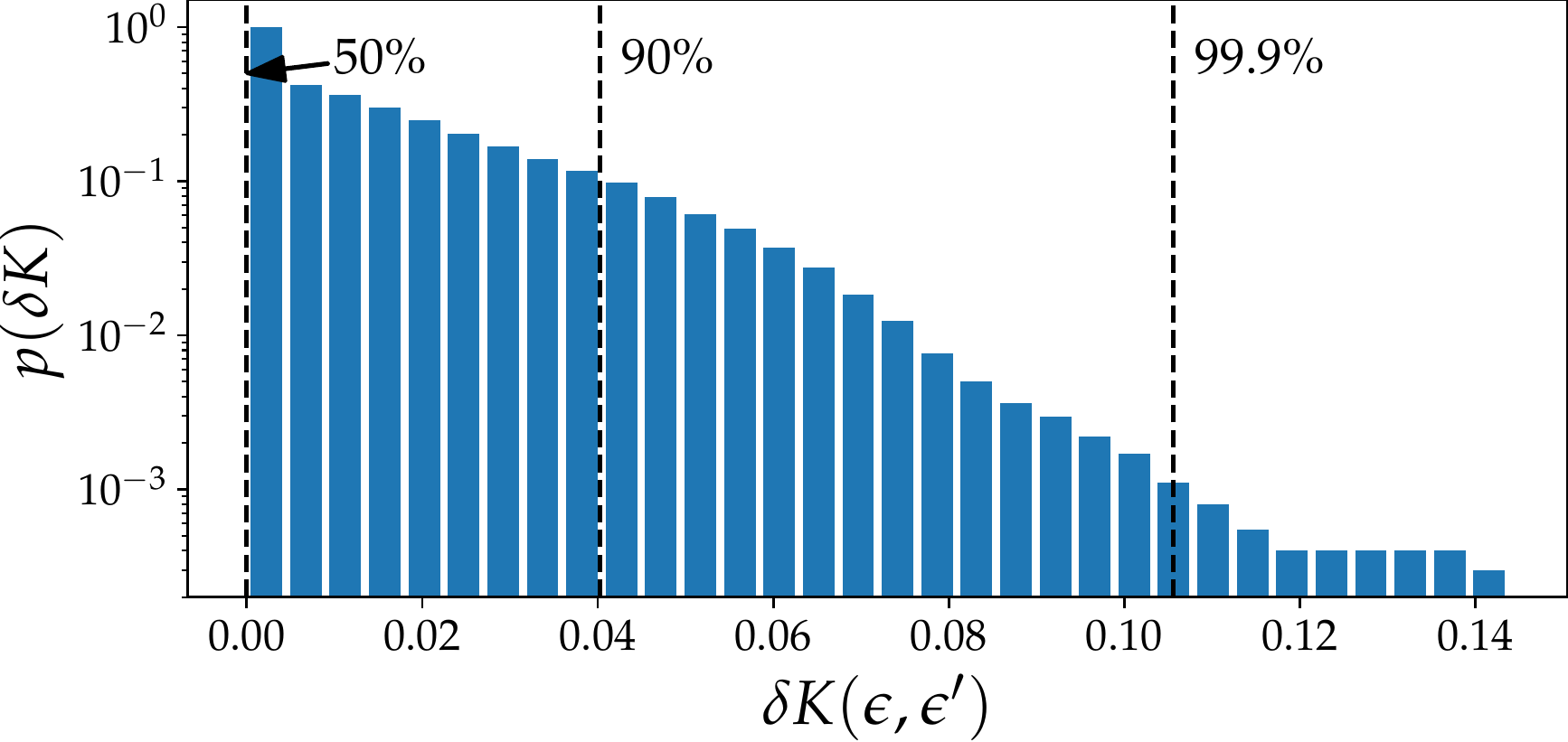}
	\caption{Cumulative distribution of the relative change of the values of the kernel in the presence of detection error (set to $\epsilon=\epsilon^{\prime}=.05$), as defined in \eqref{eq:Kij} ({\it note the logscale of the y-axis}). The average is performed over 100 estimations of the set of feature vectors, each one estimated with 10,000 measurements. The dashed vertical lines represent the 0.5, .9 and .999 quantiles.
	}
	\label{fig:noise_effect}
\end{figure}
    The histogram of the values of $\delta K_{ij}(\epsilon,\epsilon^\prime)$ obtained across 100 training sequences over the data set Fingerprint* are represented in figure \ref{fig:noise_effect}. As indicated by the arrow, for roughly half of the pairs of graphs, the relative error is below machine precision, and the distribution decays rapidly when the discrepancy increases (the relative change due to noise is above 10\% for only $\sim$0.13\% of them).
    
    This is consistent with the estimated accuracy in the noisy case, that stays close to the one of the noiseless kernel, and that still surpasses that of GS and RW (as shown in \ref{tab:pulser}).
    
\section*{Conclusion}

We proposed a new procedure for building graph kernels and analyze graph-structured data than can be implemented on quantum devices. 
This Quantum Evolution kernel is very flexible as it was illustrated on a few examples.
We showed that QE is on par with state-of-the-art graph kernels in terms of accuracy.
We were able to benchmark its expected performances on a neutral-atom quantum device, via simulation of the dynamics on small graphs, and show that the performances of the kernel are stable against detection error.
Our proposal can be readily tested on state-of-the-art neutral atom platforms. These platforms may be limited to smaller graphs (up to $\sim100$ nodes), but 
quantum-inspired algorithms have been introduced for a wide range of problems\cite{Tang18,Arrazola20}, and it has been suggested that a significant
speed-up could be gained by implementing variations of near-term quantum computing algorithms running on GPUs. 

In a short-term perspective, the most exciting follow up will be to implement a QE kernel on a real quantum machine. 

In the meantime, a lot can still be done on the theory side.
We focused on the layered time-evolution scheme, using Ising or XY Hamiltonians. But, as mentioned earlier in the text, there might be cases or platforms for which a continuous approach {\sl à la} optimized control might yield interesting results. 
We also only discussed the case of an Ising or XY graph Hamiltonian, but here as well, a lot has yet to be done in order to understand which Hamiltonian to use. More synthetic/abstract choices (using more than two -body interactions for example) could widen the range of application of this approach.
Similarly, designing the right observable to measure or the best probability distribution to construct from the dynamics of the system are also potential sources of improvement. This includes the use of multikernel learning, to combine different kernels built from the same sequence of measurements (and their associated bitstrings) on the final states.
Finally, understanding the effect of the various sources of noise on the performances of the kernel will be essential, as some of them might even contain information on the underlying graph.

\section*{Acknowledgments}
We thank Lucas Leclerc, Antoine Browaeys, Thierry Lahaye, Romain Fouilland and Christophe Jurczak for useful discussions.
It was granted access to the HPC resources of CINES under the allocation 2019-A0070911024 made by GENCI.
 It was also supported by region Ile de France through the Pack Quantique (PAQ).

\appendix
\section{Computation of the time-dependent expectation value of diagonal operators in the Ising case}
We provide here more details about analytical results in a few tractable cases. First, in the simplest case where $p=1$,
with a uniform Ising model without linear term, then in the case of the most generic Ising model, and finally in the case where 
quadratic observables are measured.
\subsection{Detailed computation in the case \texorpdfstring{$p=1$}{p=1}    \label{sec:appendixp1Ising}}
	We detail here the computations in the case of the Ising Hamiltonian with a depth of $p=1$, leading to \eqref{eq:ndensity}.
		For a given $\sigma\in\{0,1\}^N$, we note $\ket{\sigma} = \bigotimes_i\ket{\sigma_i}$, and $n_i = \Sum_{i\in\vertices} \sigma_i$, and
		we introduce the mixing operator
		\begin{align*}
			\hat{P}^\dagger_\vartheta &= \exp\left(-\ii\,\Ham_\vartheta\right)
								\equiv\Prod_i\hat{P}^\dagger_{\vartheta,i}\\
								&=\Prod_i\left[\Cos\,\Id+\Sin\left(\hat\sigma_i^+-\hat\sigma_i^-\right) \right],
		\end{align*}
		so that
		\begin{gather*}
			\hat{P}^\dagger_\vartheta\ket{\psi_0} = \Sum_{\sigma\in\{0,1\}^N} (\Cos)^{N-n_\sigma}(\Sin)^{n_\sigma}\ket{\sigma}.
		\end{gather*}
		The Ising Hamiltonian $\Ham_I$ \eqref{eq:Ising} is diagonal in the computational basis ($\Ham_I\ket{\sigma} = E_\sigma\ket{\sigma}$), therefore
		\begin{gather*}
			\e^{-\ii \Ham_I t}\hat{P}^\dagger_\vartheta\ket{\psi_0}= 
			\Sum_{\sigma\in\{0,1\}^N} (\Cos)^{N-n_\sigma}(\Sin)^{n_\sigma}\e^{-\ii E_{\sigma}}\ket{\sigma}.
		\end{gather*}
		Finally the state that is measured is
		\begin{flalign*}
			\ket{\psi_\vartheta(t)} 	&= \e^{\ii \vartheta \Ham_1}\e^{-\ii t\Ham_\Graph}\e^{-\ii \vartheta \Ham_1}\ket{\psi_0}\\
							&= \Sum_{\sigma\in\{0,1\}^N} (\Cos)^{N}(\Tan)^{n_\sigma}\e^{-\ii E_{\sigma}}			\bigotimes_i\hat{P}_{\vartheta,i}\ket{\sigma_i}
		\end{flalign*}
		The final measurement can be written as
			\begin{flalign}
				\label{eq:noft_lambda}
				n(t) = (\Cos)^{2N}\Sum_{\sigma,\sigma'\in\{0,1\}^N} \left\{
				        \e^{-\ii \left(E_\sigma-E_{\sigma'}\right)t}
				        (\Tan)^{n_\sigma+n_{\sigma'}}\right.\\\left.\times
				        \Sum_{\mu \in \{0,1\}^N} 
								n_\mu 
								\Prod_{i=1}^N
								\langle\sigma'_i|\hat{P}^\dagger_{\vartheta,i}|\mu_i\rangle
								\langle\mu_i|\hat{P}_{\vartheta,i}|\sigma_i\rangle
								\right\}.\nonumber
			\end{flalign}
		We note that
		\begin{gather*}
			\langle\mu_i|\hat{P}_{\vartheta,i}|\sigma_i\rangle =  \Cos\, \delta_{\mu_i,\sigma_i} +  \Sin\, \left(\delta_{\mu_i,\sigma_i-1} -  \delta_{\mu_i,\sigma_i+1}\right),
		\end{gather*}
		so that we get the following table
	        	\begin{gather}
	        		\label{eq:table}
	        		\begin{array}{|c|c|}\hline
	        			(\mu_i,\sigma_i,\sigma'_i)	&	\langle\sigma'_i|\hat{P}^\dagger_{\vartheta,i}|\mu_i\rangle\langle\mu_i|\hat{P}_{\vartheta,i}|\sigma_i\rangle\\\hline\hline
	        			(0,0,0),(1,1,1)				&	\Cos^2\\
	        			(0,1,1),(1,0,0)				&	\Sin^2\\
	        			(0,1,0),(0,0,1)				&	\Cos\Sin\\
	        			(1,1,0),(1,0,1)				&	-\Cos\Sin\\\hline
	        		\end{array}
	        	\end{gather}
		We can then write
		\begin{multline*}
			\Prod_{i=1}^N
				\langle\sigma'_i|\hat{P}^\dagger_{\vartheta,i}|\mu_i\rangle
				\langle\mu_i|\hat{P}_{\vartheta,i}|\sigma_i\rangle =\\
			(\Cos)^{2N}(\Tan)^{n_{\neq}+2n_1}(\Tan)^{2(n^{(0)}_{\mu}-n^{(1)}_{\mu})} (-1)^{n^{(\neq)}_{\mu}},
		\end{multline*}
		where
		\begin{align*}
			n_{1} &= \Sum_{i\in\vertices} \sigma_i\sigma'_i \equiv \left|\{ i\in\vertices | \sigma_i = \sigma'_i=1\}\right|\\
			n_{0} &= \Sum_{i\in\vertices} (1-\sigma_i)(1-\sigma'_i) \equiv \left|\{ i\in\vertices | \sigma_i = \sigma'_i=0\}\right|\\
			n_{\neq} &= \Sum_{i\in\vertices} \left[\sigma_i(1-\sigma'_i)+\sigma'_i(1-\sigma_i)\right] \equiv \left|\{ i\in\vertices | \sigma_i \neq \sigma'_i\}\right|\\
			n^{(1)}_{\mu} &= \Sum_{i\in\vertices} \sigma_i\sigma'_i\mu_i \equiv \left|\{ i\in\vertices | \sigma_i = \sigma'_i = 1\neq \mu_i\}\right|\\
			n^{(0)}_{\mu} &= \Sum_{i\in\vertices} \sigma_i\sigma'_i\mu_i \equiv \left|\{ i\in\vertices | \sigma_i = \sigma'_i = 0\neq \mu_i\}\right|\\
			n^{(\neq)}_{\mu} &= \Sum_{i\in\vertices}  \left[\sigma_i(1-\sigma'_i)+\sigma'_i(1-\sigma_i)\right]\mu_i \\
						&\equiv \left|\{ i\in\vertices | \sigma_i \neq \sigma'_i \land \mu_i = 1\}\right|.
		\end{align*}
		Therefore
		\begin{multline*}
			\Sum_{\mu \in \{0,1\}^N} n_\mu \langle\sigma'_i|\hat{P}^\dagger_{\vartheta,i}|\mu_i\rangle
				\langle\mu_i|\hat{P}_{\vartheta,i}|\sigma_i\rangle =\\
				\delta_{0,n_{\neq}}\left[n_0(\Sin)^2+n_1(\Cos)^2\right]-\delta_{1,n_{\neq}}\Cos\Sin,
		\end{multline*}
		and
		\begin{multline*}
			n(t) =2N\sin^2\vartheta\cos^2\vartheta\\-
			2(\Cos)^{2N+2}\Sum_{\sigma\in\{0,1\}^N}(\Tan)^{2n_\sigma}
			\Sum_{i|\sigma_i=1}\cos\left(\Delta_{\sigma,i}t\right),
		\end{multline*}
		where $\Delta_{\sigma,i}=E_\sigma-E_{\sigma'(i)}$ where $\sigma'(i)$ is the configuration obtained from $\sigma$ by flipping $i$.
		We can transform this sum, remarking that
		\begin{multline*}
			\Sum_{\sigma\in\{0,1\}^N}(\Tan)^{2n_\sigma}\Sum_{i|\sigma_i=1}\cos\left(\Delta_{\sigma,i}t\right)=\\
				\Sum_{i\in\vertices}\Sum_{n=0}^N (\Tan)^{2n}\Sum_{\substack{\sigma | n_\sigma=n\\ \sigma_i=1}}\cos\left(\Delta_{\sigma,i}t\right).
		\end{multline*}
		We note $\kappa_i$ the degree of node $i$. 
		For two given integers $\omega\leq \kappa_i$ and $n\geq0$, there are $\binom{N-1-\kappa_i}{n}\binom{\kappa_i}{\omega}$ permutations 
		$\sigma$ such that $n_\sigma=n+\omega+1$, $\sigma_i=1$, and $\Delta_{\sigma,i}=\omega$. From this we get
		\begin{multline*}		
			\Sum_{i\in\vertices}\Sum_{n=0}^N (\Tan)^{2n}\Sum_{\substack{\sigma | n_\sigma=n\\ \sigma_i=1}}\cos\left(\Delta_{\sigma,i}t\right)=\\
				\tan^{2}\vartheta\Sum_{i\in\vertices}\Re\left\{\Sum_{\omega=0}^{\kappa_i}\binom{\kappa_i}{\omega}\left(\e^{\ii t}\tan^{2}\vartheta\right)^\omega\right\}\times\\
				 \left[\Sum_{n=0}^{N-1-\kappa_i}\binom{N-1-\kappa_i}{n}(\tan^{2}\vartheta)^{n}\right],
		\end{multline*}
		and
		\begin{multline*}		
			\Sum_{i\in\vertices}\Sum_{n=0}^N (\Tan)^{2n}\Sum_{\substack{\sigma | n_\sigma=n\\ \sigma_i=1}}\cos\left(\Delta_{\sigma,i}t\right)=\\
				\tan^{2}\vartheta\Sum_{i\in\vertices}\Re\left\{\left(1+\e^{\ii t}\tan^{2}\vartheta\right)^{\kappa_i}\right\}\times\\
				 \left(1+\tan^{2}\vartheta)^{N-1-\kappa_i}\right],
		\end{multline*}
		The term in the sum over $i$ only depends on $\kappa_i$ so, if we note $m_\Graph(\kappa)$ the number of vertices
		 of degree $\kappa$ in $\Graph$, we obtain \eqref{eq:ndensity}.

\subsection{Extensions to quadratic observables}
Using the same setup, but measuring the expectation value of the Ising Hamiltonian instead of the occupation, further insight on the graph can be obtained. In particular, correlation functions appear explicitely. The expecation value of the observable is given by
\begin{align*}
	\braket{\Ham_I(t)} =& \Sum_{(i,j)\in\edges} \braket{\hat{n}_i\hat{n}_j}(t)\\
				= & \;4\sin^2\vartheta\cos^2\vartheta\left|\edges\right|+\\
				& \; 2\sin^2\vartheta(\Cos)^{2N+2}\Sum_{\vertices'\subset\vertices}\Sum_{(i,j)\in\edges}
				\Real{f_{n'_i,n'_j}(t)},
\end{align*}
where the sum runs over all subgraphs $\vertices'\subset \vertices$, $n'_i=\left|\left\{k\in\vertices'|(i,k)\in\edges\right\}\right|$ is the number of neighbors of $i$ in $\vertices'$, and
\begin{align*}
    f_{n,m}(t) =& \;\e^{\ii(1+n+m)t}w(t)^2+\e^{\ii(n-m)t}|w(t)|^2\\
                &    -2(\e^{\ii nt}+\e^{\ii mt})w(t),
\end{align*}
with $w(t) = \left(1 + \tan^2\vartheta\;\e^{i t}\right)^{-1}$.

\subsection{Extension to a generic Ising model}
We can extend the computation presented in \ref{sec:Ising} to the case of a graph with labeled vertices and/or
weighted edges. For a given graph $\Graph = (\vertices,\edges)$,
with vertex labels $\{h_i\}_{i\in\vertices}$ and edge labels $\{J_{ij}\}_{(i,j)\in\edges}$
\footnote{For any $(i,j)\in\vertices^2$ such that $(i,j)\notin\edges$, we set $J_{ij}=0$.},
let us consider a set of $N=|\vertices|$ qubits interacting with the Hamiltonian
\begin{gather}
	\Ham_\Graph=\Sum_{(i,j)\in \mathcal{E}} J_{ij}\hat\sigma^z_{i} \hat\sigma^z_{j} + \Sum_{i\in \vertices} h_i\hat\sigma^z_i.
\end{gather}
In this case, the procedure described in \ref{sec:Ising} leads to a more generic version of \eqref{eq:ndensity} :

\begin{widetext}
\begin{align}
	\label{eq:ngeneric}
n(t)&=2\sin^2\vartheta\cos^2\vartheta\,\,\Sum_{i\in\vertices}\Re \left\{1-
		\e^{\ii h_i t}\Prod_{j\in\vertices}
		\left({\cos^2\vartheta+\sin^2\vartheta\;\e^{\ii J_{ij}t}}\right)
		\right\}\\
	&=2N\sin^2\vartheta\cos^2\vartheta+		
		2\Sum_{\tilde{\vertices}\subset\vertices}
		(\Cos)^{2(N-|\tilde{\vertices}|+1)}(\Sin)^{2|\tilde{\vertices}|+2}
		\Sum_{i\in\vertices} 
		\cos{\left[\left(h_i+\Sum_{j\in\tilde{\vertices}}J_{ij}\right)t\right]}.
\end{align}
For a given $i\in\vertices$, we note ${\vertices}_i^1 = \left\{j\in{\vertices}|(i,j)\in\edges\right\}$ the set of neighbors of $i$, and ${\vertices}_i^0 = \left\{j\in{\vertices}|(i,j)\notin\edges\right\}=\vertices\setminus\vertices_i^1$ its complement in $\vertices$, such that
\begin{align*}
n(t)
	&=2N\sin^2\vartheta\cos^2\vartheta+		
		2\cos^{2}\vartheta\sin^2\vartheta\Sum_{i\in\vertices} 
		\cos^{2\kappa_i}\vartheta
		\Sum_{\tilde{\vertices}^1\subset\vertices_i^1}(\Tan)^{2|\tilde{\vertices}^1|}
		\cos{\left[\left(h_i+\Sum_{j\in\tilde{\vertices^1}}J_{ij}\right)t\right]}.
\end{align*}
\end{widetext}
The Fourier transform would then have peaks at $k = h_i+\Sum_{j\in\tilde\vertices_i}J_{ij}$,
where $\tilde\vertices_i\subset \vertices^1_i$ is any subset of the
neighbourhood of $i$, with weight $\propto\left[(\Sin)^{|\tilde{\vertices}^1|}(\Cos)^{\kappa_i-|\tilde{\vertices}^1|}\right]^2$

\subsection{Quadratic observables}
Even with an Ising graph Hamiltonian, further information can be reached. We consider here the graph Hamiltonian to be the most generic Ising Hamiltonian 
\begin{gather}
    \Ham_\Graph = \Sum_{i<j} J_{ij}\hat n_i \hat n_j -\Sum_i h_i \hat n_i
\end{gather}

By measuring quadratic (and higher order) observables, one can extract information about the subgraphs of the system. For example, if one were to measure the {\it pure} Ising energy, ({\sl i.e.} $\Obs = \sum_{(i,j)\in\edges} \hat n_i\hat n_j$)
\begin{widetext}
	\begin{multline}
\langle n_{v_1}n_{v_2}\rangle(t) =4\sin^4\vartheta\cos^4\vartheta\,\,\Re \left\{1
		-\e^{\ii h_{v_1} t}w_{v_1v_2}(t)^{-1}\Prod_{ v\in\vertices}w_{v_1 v}(t)
		-\e^{\ii h_{v_2} t}w_{v_1v_2}(t)^{-1}\Prod_{ v\in\vertices}w_{v_2 v}(t)
		\right.\\
		+\frac{1}{2}\e^{\ii (h_{v_1}+h_{v_2} +J_{v_1v_2}) t}w_{v_1v_2}(t)^{-2}
		    \Prod_{ v\in\vertices}w^+_{ v}(t)
		\left.+\frac{1}{2}\e^{\ii (h_{v_1}-h_{v_2}) t}|w_{v_1v_2}(t)|^{-2}
		    \Prod_{ v\in\vertices}w^-_{ v}(t),
\right\}\nonumber
	\end{multline}
\end{widetext}
where
\begin{gather}
    \left\{
    \begin{array}{lcl}
    w_{vv'}(t) &=& {\cos^2\vartheta+\sin^2\vartheta\;\e^{\ii J_{v v'}t}}\\
    w_{v}^{\pm}(t) &=& {\cos^2\vartheta+\sin^2\vartheta\;\e^{\ii (J_{v_1 v}\pm J_{v_2v})t}}
    \end{array}
    \right.
\end{gather}

\section{Classical graph kernels}
\label{sec:appendix classical kernels}

We briefly introduce here the classical kernels used for benchmark comparison, the Random Walk kernel \cite{Vishwanathan10} and the Graphlet Subsampling \cite{Przulj07, pmlr-v5-shervashidze09a} kernel.

\subsection{Random Walk kernel}

The random walk kernel aims at counting the number of random walks of different sizes between two graphs. Given $\Graph = (\vertices, \edges)$ and $\Graph' = (\vertices', \edges')$ we define the product graph $\Graph_\times$ as the graph with the set of vertices $\vertices_\times = \{(v, v') \in \vertices \times \vertices' \}$ and the set of edges $\edges_\times = \{(e, e') \in \edges \times \edges' \}$. We denote then $A_\times$ the adjacency matrix of the product graph. The geometric random walk kernel for a positive weight $\lambda$ is defined as 
\begin{gather}
    k(\Graph, \Graph') = \Sum_{p,q=1}^{|V_\times|} \Big[\Sum_{l=0}^{\infty} \lambda^l A_\times^l \Big]_{pq} = e^T(I - \lambda A_\times)^{-1}e,
\end{gather}
where $e$ is the vector full of 1s and $I$ is the identity matrix. This kernel is only defined when $\lambda$ is smaller than the inverse of the biggest eigenvalue of $A_\times$ in order to allow convergence. This implies that small values of $\lambda$ are tested in practice, which makes the contribution of long random walks vanish and limits the ability to capture complex features. 

\subsection{Graphlet Subsampling kernel}

Let $G_k$ be the set of all possible graphlets of size $k$. For a given graph $\Graph$, we define $f_k(\Graph)$ the vector such that the i-th component is the number of graphlets isomorphic to the i-th element of $G_k$ normalized by the the total number of graphlets of $\Graph$. The kernel is then defined by 
\begin{gather}
    k(\Graph, \Graph') = f_k(\Graph)^T f_k(\Graph')
\end{gather}

An exact evaluation of this kernel requires an exhaustive enumeration of the graphlets of each graph, which requires $\mathcal{O}(n^k)$ operations since there are $\binom{n}{k}$ graphlets in $\Graph$, $n$ being the number of vertices. We resort then to randomly sampling a fixed number of graphlets in the graph, which leads to a good approximation (as described in \cite{pmlr-v5-shervashidze09a}).

\section{Bayesian Optimization\label{app:bayes}}

We are interested in the problem of finding the maximum of a function $f: \mathcal{X} \longrightarrow \mathbb{R}$ where $\mathcal{X}$ is a space of parameters, and $f$ is a costly function to evaluate. In our case $\mathcal{X}$ would be the space of parameters $\boldsymbol{\Lambda}$, and $f$ would be the mean of the accuracy score in a cross-validation on a particular dataset. One way to do so is the bayesian optimization, which we will briefly summarize in this section \cite{frazier2018tutorial}.

Since the function $f$ is very costly to evaluate, we build a model that approximates it. This model is called a surrogate function noted $\Tilde{f}$, and is usually a gaussian process. A gaussian process $\mathcal{GP}$ is completely defined by a mean function $\mu(x)$ and a covariance kernel $k(x, x')$ such that for any collection of points $(x_1, \dots, x_n)$, the vector $[\mathcal{GP}(x_1), \dots \mathcal{GP}(x_n)]$ follows a multivariate normal of means $[\mu(x_1), \dots \mu(x_n)]$ and covariance matrix given by the entries $k(x_i, x_j)$.

The kernel used is the Matérn kernel\cite{Matern60} given by
\begin{equation}
    k(x, x') = \alpha_0 \frac{2^{1-\nu}}{\Gamma(\nu)}(\sqrt{2\nu}\|x-x'\|)^\nu K_\nu(\sqrt{2\nu}\|x-x'\|)
\end{equation}
where $\Gamma$ is the gamma function, $K_\nu$ is the modified Bessel function, and $\|x-x'\|^2 = \Sum_i \alpha_i (x_i-x'_i)^2$ is a weighted norm.

For the multikernel learning, the kernel used is the Gaussian Radial Basis kernel:
\begin{equation}
    k(x, x') = \alpha_0 \exp(\|x-x'\|^2)
\end{equation}
where $\|x-x'\|^2 = \Sum_i \alpha_i (x_i-x'_i)^2$ is a weighted norm.

Given some observations of $f$, the objective is to find the best surrogate function $\Tilde{f}$. One then has to define an acquisition function $\alpha(x)$ that will give the information on where to perform the next evaluation of $f$. One commonly used acquisition function is the lower confidence bound defined by $\alpha(x) = \mu(x) + \kappa \sigma(x)$. $\kappa$ is a positive parameters representing the trade-off between exploitation and exploration. 

Instead of sequentially evaluating the function and updating the gaussian process, one can use a parallelised version of the algorithm \cite{kandasamy2018parallelised}. Instead of minimizing an acquisition function, the next point to evaluate is determined by the minimum of a sample of the gaussian process. The surrogate function is then updated by batches.
\SetInd{2.em}{0.1em}
\begin{algorithm}[htpb]
\SetAlgoLined
 Observe $f$ at $n_0$ points\;
 Fit the gaussian process $\Tilde{f}$\;
 \For{$n_0 \:\leq\: i \:\leq \:N$}{
  Find $x^*_i = \text{argmin}_x \alpha(x)\;$\\
  Evaluate $f$ at $x^*_i$\;
  Update $\Tilde{f}$ with the new couple $x^*_i, f(x^*_i)$
 }
 \KwResult{$\text{argmin} (f(x^*_1), \dots, f(x^*_N))$}
 \caption{Bayesian optimization with gaussian surrogate}
\end{algorithm}

\begin{algorithm}[htpb]
\SetAlgoLined
 Observe $f$ at $n_0$ points\;
 Fit the gaussian process $\Tilde{f}$\;
 \For{$n_0 \:\leq\: i \:\leq \:N$}{
     \For{each worker j}{
      Sample $\hat{f}$ from $\Tilde{f}\;$\\
      Find $x^*_j = \text{argmin}_x \hat{f}(x)\;$\\
      Evaluate $f$ at $x^*_j$\;
     }
  Aggregate the values $x^*_j, f(x^*_j)$ for each worker\;
  Update $\Tilde{f}$\; 
 }
 \KwResult{$\text{argmin} (f(x^*_1), \dots f(x^*_{kN}))$}
 \caption{Parallel bayesian optimization with gaussian surrogate and $k$ workers}
\end{algorithm}

\section{Multiple kernel learning \label{app:Multikernel}}

We explore here a postprocessing strategy to improve the score. Multikernel learning is the idea of combining several kernels to construct a better one \cite{gonen2011multiple}. Such a combination can be made in several way, the most common ones being by linear combination or product. Given a family of kernels $(K_1, \dots, K_R)$, we are interested in building the new kernel

\begin{equation}
    K_{{\bf p}} = \Sum_{i=1}^R p_i K_i,
\end{equation}

where ${\bf p} = (p_1, \dots, p_R)$ is a R-uplet of non-negative numbers. For a fixed $\boldsymbol{p}$ a new kernel is created, and a model can be trained on it. The question is whether there exists a ${\bf p}$ such that the model given by $K_{{\bf p}}$ is better than any model given by the individuals kernels.

In our case, a kernel defined for each parameter $\Lambda$. $R$ values of the parameters are chosen and an optimal combination of their corresponding kernels is determined. The entries of ${\bf p}$ are bounded between 0 and 1, and the best value is found again by bayesian optimization. This method uses the results obtained previously by the quantum evolution, and can be done classically without any further consumption of quantum resources.

\begin{figure}[ht!]
\includegraphics[width=\linewidth]{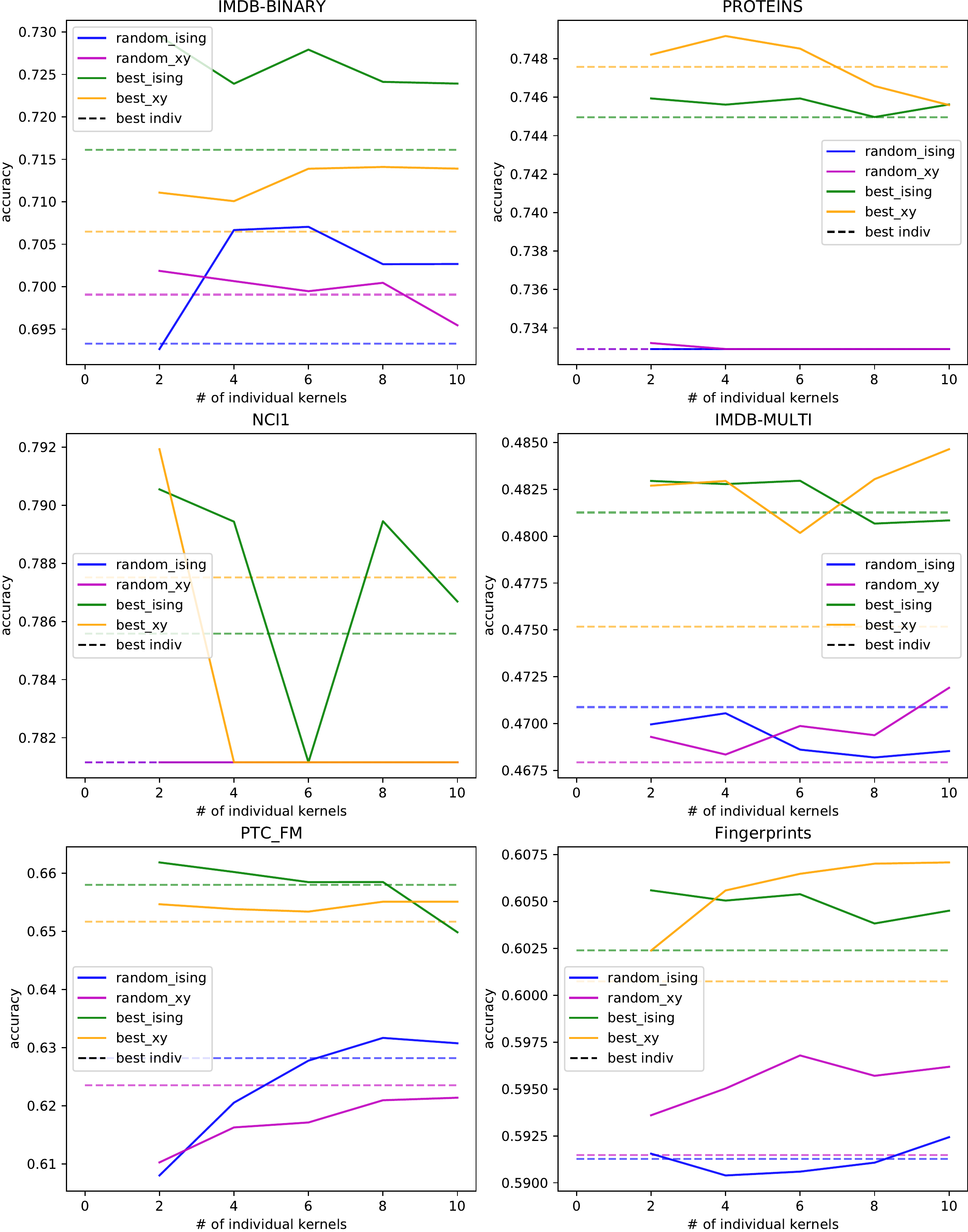}
\caption{Accuracy obtained after multiple kernel learning as a function of the number of individual kernels. Each panel corresponds to a different dataset. In each one, the solid green (yellow) curve is obtained by combining the $R$ best performing kernels in the Ising ($XY$) scheme, and the solid blue (purple) is obtained by combining $R$ random kernels in the Ising ($XY$) scheme. For each of these curves, a doted line of the same color represents the accuracy of the best performing kernel among the ones that are combined.}
\label{fig:multi_kernel}
\end{figure}

The procedure is tested in different settings. For each Hamiltonian type Ising or XY, up to $R=10$ different kernels are used, as described in \ref{sec:Benchmarking}, Two ways of selecting these kernels are tested. In the first one, $R$  kernels are chosen randomly while in the second approach, only the $R$ best performing kernels are considered. The optimization is performed by making $50\times n_{kernel}$ calls and $20\times n_{kernel}$ initial evaluations where $n_{kernel}$ is the number of individual kernels to consider. The results are shown in figure \ref{fig:multi_kernel}. The procedure is most of the time successful --- \emph{i.e} a combination of individual kernels manages to get a better score than the best of single kernels ---, and the score improves with the number of kernels combined, as one would expect theoretically. However, because of the limitations of the optimization procedures, it is not always the case here. For IMDB-BINARY, PROTEINS, Fingerprints a better score is consistently reached for almost every configuration. For NCI1, only the 2 best XY kernels and the best Ising kernels can be combined to produce a better score. Random Ising kernels for IMDB-MULTI and random kernels for PTC\_FM also fail. It is interesting to note that the best combination of random kernels never beats the best individual kernel with optimized parameters, except in the case of IMDB-BINARY where the best combination of 4 and 6 random Ising kernels beats the best XY kernel. 

\bibliographystyle{unsrtnat}
\bibliography{refs}

\begin{thebibliography}{62}
\providecommand{\natexlab}[1]{#1}
\providecommand{\url}[1]{\texttt{#1}}
\expandafter\ifx\csname urlstyle\endcsname\relax
  \providecommand{\doi}[1]{doi: #1}\else
  \providecommand{\doi}{doi: \begingroup \urlstyle{rm}\Url}\fi

\bibitem[Varnek and Baskin(2012)]{Varnek2012}
Alexandre Varnek and Igor Baskin.
\newblock Machine learning methods for property prediction in chemoinformatics:
  Quo vadis?
\newblock \emph{Journal of Chemical Information and Modeling}, 52\penalty0
  (6):\penalty0 1413--1437, 2012.
\newblock \doi{10.1021/ci200409x}.
\newblock URL \url{https://doi.org/10.1021/ci200409x}.
\newblock PMID: 22582859.

\bibitem[Gilmer et~al.(2017)Gilmer, Schoenholz, Riley, Vinyals, and
  Dahl]{gilmer17}
Justin Gilmer, Samuel~S. Schoenholz, Patrick~F. Riley, Oriol Vinyals, and
  George~E. Dahl.
\newblock Neural message passing for quantum chemistry.
\newblock In Doina Precup and Yee~Whye Teh, editors, \emph{Proceedings of the
  34th International Conference on Machine Learning}, volume~70 of
  \emph{Proceedings of Machine Learning Research}, pages 1263--1272. PMLR,
  06--11 Aug 2017.
\newblock URL \url{http://proceedings.mlr.press/v70/gilmer17a.html}.

\bibitem[Muzio et~al.(2020)Muzio, O’Bray, and Borgwardt]{Muzio20}
Giulia Muzio, Leslie O’Bray, and Karsten Borgwardt.
\newblock {Biological network analysis with deep learning}.
\newblock \emph{Briefings in Bioinformatics}, 22\penalty0 (2):\penalty0
  1515--1530, 11 2020.
\newblock ISSN 1477-4054.
\newblock \doi{10.1093/bib/bbaa257}.
\newblock URL \url{https://doi.org/10.1093/bib/bbaa257}.

\bibitem[Borgwardt et~al.(2005)Borgwardt, Ong, Schönauer, Vishwanathan, Smola,
  and Kriegel]{Borgwardt05}
K.~M. Borgwardt, C.~S. Ong, S.~Schönauer, S.~V. Vishwanathan, A.~J. Smola, and
  H.~P. Kriegel.
\newblock Protein function prediction via graph kernels.
\newblock \emph{Bioinformatics}, 21, 2005.
\newblock \doi{10.1093/bioinformatics/bti1007}.

\bibitem[Scott(2011)]{Scott11}
John Scott.
\newblock {Social network analysis: developments, advances, and prospects}.
\newblock \emph{Social Network Analysis and Mining}, 1\penalty0 (1), 01 2011.
\newblock \doi{10.1007/s13278-010-0012-6}.
\newblock URL \url{https://doi.org/10.1007/s13278-010-0012-6}.

\bibitem[Harchaoui and Bach(2007)]{harchaoui2007image}
Za{\"\i}d Harchaoui and Francis Bach.
\newblock Image classification with segmentation graph kernels.
\newblock In \emph{2007 IEEE Conference on Computer Vision and Pattern
  Recognition}, pages 1--8. IEEE, 2007.

\bibitem[Nikolentzos et~al.(2017)Nikolentzos, Meladianos, Rousseau, Stavrakas,
  and Vazirgiannis]{nikolentzos2017shortest}
Giannis Nikolentzos, Polykarpos Meladianos, Fran{\c{c}}ois Rousseau, Yannis
  Stavrakas, and Michalis Vazirgiannis.
\newblock Shortest-path graph kernels for document similarity.
\newblock In \emph{Proceedings of the 2017 Conference on Empirical Methods in
  Natural Language Processing}, pages 1890--1900, 2017.

\bibitem[Glava{\v{s}} and {\v{S}}najder(2013)]{glavavs2013recognizing}
Goran Glava{\v{s}} and Jan {\v{S}}najder.
\newblock Recognizing identical events with graph kernels.
\newblock In \emph{Proceedings of the 51st Annual Meeting of the Association
  for Computational Linguistics (Volume 2: Short Papers)}, pages 797--803,
  2013.

\bibitem[Latouche and Rossi(2015)]{Latouche15}
Pierre Latouche and Fabrice Rossi.
\newblock Graphs in machine learning: an introduction.
\newblock \emph{European Symposium on Artificial Neural Networks, Computational
  Intelligence and Machine Learning (ESANN)}, pages 207--218, April 2015.

\bibitem[Kriege et~al.(2020)Kriege, Johansson, and Morris]{Kriege20}
Niels~M. Kriege, Fredrik~D. Johansson, and Christopher Morris.
\newblock A survey on graph kernels.
\newblock \emph{Applied Network Science}, 5\penalty0 (6), 2020.
\newblock \doi{10.1007/s41109-019-0195-3}.
\newblock URL \url{https://doi.org/10.1007/s41109-019-0195-3}.

\bibitem[Borgwardt et~al.(2020)Borgwardt, Ghisu, Llinares-López, O'Bray, and
  Rieck]{Borgwardt20}
Karsten~M. Borgwardt, M.~Elisabetta Ghisu, Felipe Llinares-López, Leslie
  O'Bray, and Bastian Rieck.
\newblock Graph kernels: State-of-the-art and future challenges.
\newblock \emph{Foundations and Trends in Machine Learning}, 13\penalty0 (5-6),
  2020.
\newblock \doi{10.1561/2200000076}.
\newblock URL \url{https://doi.org/10.1561/2200000076}.

\bibitem[Schlk{\"o}pf and Smola(2001)]{Schlkopf01}
B.~Schlk{\"o}pf and A.~J. Smola.
\newblock \emph{Learning with Kernels: Support Vector Machines, Regularization,
  Optimization, and Beyond}.
\newblock The MIT Press, 2001.

\bibitem[Schuld and Killoran(2019)]{Schuld19}
Maria Schuld and Nathan Killoran.
\newblock Quantum machine learning in feature hilbert spaces.
\newblock \emph{Phys. Rev. Lett.}, 122:\penalty0 040504, Feb 2019.
\newblock \doi{10.1103/PhysRevLett.122.040504}.
\newblock URL \url{https://link.aps.org/doi/10.1103/PhysRevLett.122.040504}.

\bibitem[Havl{\'\i}{\v c}ek et~al.(2019)Havl{\'\i}{\v c}ek, C{\'o}rcoles,
  Temme, Harrow, Kandala, Chow, and Gambetta]{Havlek19}
Vojt{\v e}ch Havl{\'\i}{\v c}ek, Antonio~D. C{\'o}rcoles, Kristan Temme,
  Aram~W. Harrow, Abhinav Kandala, Jerry~M. Chow, and Jay~M. Gambetta.
\newblock Supervised learning with quantum-enhanced feature spaces.
\newblock \emph{Nature}, 567\penalty0 (7747):\penalty0 209--212, 2019.
\newblock \doi{10.1038/s41586-019-0980-2}.
\newblock URL \url{https://doi.org/10.1038/s41586-019-0980-2}.

\bibitem[Kishi et~al.(2021)Kishi, Satoh, Raymond, Yamamoto, and
  Sakakibara]{kishi21}
Kaito Kishi, Takahiko Satoh, Rudy Raymond, Naoki Yamamoto, and Yasubumi
  Sakakibara.
\newblock Graph kernels encoding features of all subgraphs by quantum
  superposition, 2021.

\bibitem[{Verdon} et~al.(2019){Verdon}, {McCourt}, {Luzhnica}, {Singh},
  {Leichenauer}, and {Hidary}]{Verdon19}
Guillaume {Verdon}, Trevor {McCourt}, Enxhell {Luzhnica}, Vikash {Singh},
  Stefan {Leichenauer}, and Jack {Hidary}.
\newblock {Quantum Graph Neural Networks}.
\newblock \emph{arXiv e-prints}, art. arXiv:1909.12264, September 2019.

\bibitem[Rossi et~al.(2015)Rossi, Torsello, and Hancock]{Rossi15}
Luca Rossi, Andrea Torsello, and Edwin Hancock.
\newblock Measuring graph similarity through continuous-time quantum walks and
  the quantum jensen-shannon divergence.
\newblock \emph{Physical Review E}, 91:\penalty0 022815, 02 2015.
\newblock \doi{10.1103/PhysRevE.91.022815}.

\bibitem[Bai et~al.(2015)Bai, Rossi, Ren, Zhang, and Hancock]{Bai15}
Lu~Bai, Luca Rossi, Peng Ren, Zhihong Zhang, and Edwin~R. Hancock.
\newblock {A Quantum Jensen-Shannon Graph Kernel Using Discrete-Time Quantum
  Walks}.
\newblock In Cheng-Lin Liu, Bin Luo, Walter~G. Kropatsch, and Jian Cheng,
  editors, \emph{Graph-Based Representations in Pattern Recognition}, pages
  252--261, Cham, 2015. Springer International Publishing.
\newblock ISBN 978-3-319-18224-7.
\newblock URL \url{https://doi.org/10.1007/978-3-319-18224-7_25}.

\bibitem[Schuld et~al.(2020)Schuld, Br\'adler, Israel, Su, and Gupt]{Schuld20}
Maria Schuld, Kamil Br\'adler, Robert Israel, Daiqin Su, and Brajesh Gupt.
\newblock Measuring the similarity of graphs with a gaussian boson sampler.
\newblock \emph{Phys. Rev. A}, 101:\penalty0 032314, Mar 2020.
\newblock \doi{10.1103/PhysRevA.101.032314}.
\newblock URL \url{https://link.aps.org/doi/10.1103/PhysRevA.101.032314}.

\bibitem[{Farhi} et~al.(2014){Farhi}, {Goldstone}, and {Gutmann}]{Farhi14}
Edward {Farhi}, Jeffrey {Goldstone}, and Sam {Gutmann}.
\newblock {A Quantum Approximate Optimization Algorithm}.
\newblock \emph{arXiv e-prints}, art. arXiv:1411.4028, Nov 2014.

\bibitem[Werschnik and Gross(2007)]{Werschnik07}
J~Werschnik and E~K~U Gross.
\newblock Quantum optimal control theory.
\newblock \emph{Journal of Physics B: Atomic, Molecular and Optical Physics},
  40\penalty0 (18):\penalty0 R175--R211, sep 2007.
\newblock \doi{10.1088/0953-4075/40/18/r01}.
\newblock URL \url{https://doi.org/10.1088/0953-4075/40/18/r01}.

\bibitem[Weimer et~al.(2010)Weimer, Müller, Zoller, and Büchler]{Weimer10}
Hendrik Weimer, Igor Müller, MarkusLesanovsky, Peter Zoller, and Hans~Peter
  Büchler.
\newblock A rydberg quantum simulator.
\newblock \emph{Nature Physics}, 6:\penalty0 382–388, Mar 2010.
\newblock \doi{10.1038/nphys1614}.
\newblock URL \url{https://doi.org/10.1038/nphys1614}.

\bibitem[{Bernien} et~al.(2017){Bernien}, {Schwartz}, {Keesling}, {Levine},
  {Omran}, {Pichler}, {Choi}, {Zibrov}, {Endres}, {Greiner}, {Vuleti{\'c}}, and
  {Lukin}]{Bernien17}
Hannes {Bernien}, Sylvain {Schwartz}, Alexander {Keesling}, Harry {Levine},
  Ahmed {Omran}, Hannes {Pichler}, Soonwon {Choi}, Alexander~S. {Zibrov},
  Manuel {Endres}, Markus {Greiner}, Vladan {Vuleti{\'c}}, and Mikhail~D.
  {Lukin}.
\newblock {Probing many-body dynamics on a 51-atom quantum simulator}.
\newblock \emph{Nature}, 551\penalty0 (7682):\penalty0 579--584, November 2017.
\newblock \doi{10.1038/nature24622}.

\bibitem[{Labuhn} et~al.(2016){Labuhn}, {Barredo}, {Ravets}, {de
  L{\'e}s{\'e}leuc}, {Macr{\`\i}}, {Lahaye}, and {Browaeys}]{Labuhn16}
Henning {Labuhn}, Daniel {Barredo}, Sylvain {Ravets}, Sylvain {de
  L{\'e}s{\'e}leuc}, Tommaso {Macr{\`\i}}, Thierry {Lahaye}, and Antoine
  {Browaeys}.
\newblock {Tunable two-dimensional arrays of single Rydberg atoms for realizing
  quantum Ising models}.
\newblock \emph{Nature}, 534\penalty0 (7609):\penalty0 667--670, Jun 2016.
\newblock \doi{10.1038/nature18274}.

\bibitem[Nikolentzos et~al.(2019)Nikolentzos, Siglidis, and
  Vazirgiannis]{nikolentzos19}
Giannis Nikolentzos, Giannis Siglidis, and Michalis Vazirgiannis.
\newblock Graph kernels: A survey, 2019.

\bibitem[Bishop(2006)]{Bishop06}
C.~Bishop.
\newblock \emph{{Pattern Recognition and Machine Learning}}.
\newblock {Springer-Verlag New York}, Cambridge, 2006.

\bibitem[Morris et~al.(2020)Morris, Kriege, Bause, Kersting, Mutzel, and
  Neumann]{Morris+2020}
Christopher Morris, Nils~M. Kriege, Franka Bause, Kristian Kersting, Petra
  Mutzel, and Marion Neumann.
\newblock Tudataset: A collection of benchmark datasets for learning with
  graphs.
\newblock In \emph{ICML 2020 Workshop on Graph Representation Learning and
  Beyond (GRL+ 2020)}, 2020.
\newblock URL \url{www.graphlearning.io}.

\bibitem[Pawara et~al.(2020)Pawara, Okafor, Groefsema, He, Schomaker, and
  Wiering]{Pawara20}
Pornntiwa Pawara, Emmanuel Okafor, Marc Groefsema, Sheng He, Lambert~R.B.
  Schomaker, and Marco~A. Wiering.
\newblock One-vs-one classification for deep neural networks.
\newblock \emph{Pattern Recognition}, 108:\penalty0 107528, 2020.
\newblock ISSN 0031-3203.
\newblock \doi{https://doi.org/10.1016/j.patcog.2020.107528}.
\newblock URL
  \url{https://www.sciencedirect.com/science/article/pii/S0031320320303319}.

\bibitem[Lin(1991)]{Lin91}
J.~Lin.
\newblock Divergence measures based on the shannon entropy.
\newblock \emph{IEEE Transactions on Information Theory}, 37\penalty0
  (1):\penalty0 145--151, 1991.
\newblock \doi{10.1109/18.61115}.

\bibitem[Erdős and Rényi(1959)]{Erdos59}
P.~Erdős and A.~Rényi.
\newblock On random graphs. i.
\newblock \emph{Publicationes Mathematicae}, 6, 1959.

\bibitem[Shervashidze et~al.(2011)Shervashidze, Schweitzer, van Leeuwen,
  Mehlhorn, and Borgwardt]{Shervashidze11}
Nino Shervashidze, Pascal Schweitzer, Erik~Jan van Leeuwen, Kurt Mehlhorn, and
  Karsten~M. Borgwardt.
\newblock Weisfeiler-lehman graph kernels.
\newblock \emph{Journal of Machine Learning Research}, 12\penalty0
  (77):\penalty0 2539--2561, 2011.
\newblock URL \url{http://jmlr.org/papers/v12/shervashidze11a.html}.

\bibitem[Bai(2014)]{Bai_thesis}
Lu~Bai.
\newblock \emph{{Information Theoretic Graph Kernels}}.
\newblock Theses, {Department of Computer Science, University of York}, May
  2014.

\bibitem[Childs et~al.(2003)Childs, Cleve, Deotto, Farhi, Gutmann, and
  Spielman]{Childs03}
Andrew~M. Childs, Richard Cleve, Enrico Deotto, Edward Farhi, Sam Gutmann, and
  Daniel~A. Spielman.
\newblock Exponential algorithmic speedup by a quantum walk.
\newblock In \emph{Proceedings of the Thirty-Fifth Annual ACM Symposium on
  Theory of Computing}, STOC '03, page 59–68, New York, NY, USA, 2003.
  Association for Computing Machinery.
\newblock ISBN 1581136749.
\newblock \doi{10.1145/780542.780552}.
\newblock URL \url{https://doi.org/10.1145/780542.780552}.

\bibitem[Pržulj(2007)]{Przulj07}
Nataša Pržulj.
\newblock {Biological network comparison using graphlet degree distribution}.
\newblock \emph{Bioinformatics}, 23\penalty0 (2):\penalty0 e177--e183, 01 2007.
\newblock ISSN 1367-4803.
\newblock \doi{10.1093/bioinformatics/btl301}.
\newblock URL \url{https://doi.org/10.1093/bioinformatics/btl301}.

\bibitem[Shervashidze et~al.(2009)Shervashidze, Vishwanathan, Petri, Mehlhorn,
  and Borgwardt]{pmlr-v5-shervashidze09a}
Nino Shervashidze, SVN Vishwanathan, Tobias Petri, Kurt Mehlhorn, and Karsten
  Borgwardt.
\newblock Efficient graphlet kernels for large graph comparison.
\newblock In David van Dyk and Max Welling, editors, \emph{Proceedings of the
  Twelth International Conference on Artificial Intelligence and Statistics},
  volume~5 of \emph{Proceedings of Machine Learning Research}, pages 488--495,
  Hilton Clearwater Beach Resort, Clearwater Beach, Florida USA, 16--18 Apr
  2009. PMLR.
\newblock URL \url{http://proceedings.mlr.press/v5/shervashidze09a.html}.

\bibitem[Vishwanathan et~al.(2010)Vishwanathan, Schraudolph, Kondor, and
  Borgwardt]{Vishwanathan10}
S.V.N. Vishwanathan, Nicol~N. Schraudolph, Risi Kondor, and Karsten~M.
  Borgwardt.
\newblock Graph kernels.
\newblock \emph{Journal of Machine Learning Research}, 11\penalty0
  (40):\penalty0 1201--1242, 2010.
\newblock URL \url{http://jmlr.org/papers/v11/vishwanathan10a.html}.

\bibitem[Siglidis et~al.(2020)Siglidis, Nikolentzos, Limnios, Giatsidis,
  Skianis, and Vazirgiannis]{JMLR:v21:18-370}
Giannis Siglidis, Giannis Nikolentzos, Stratis Limnios, Christos Giatsidis,
  Konstantinos Skianis, and Michalis Vazirgiannis.
\newblock Grakel: A graph kernel library in python.
\newblock \emph{Journal of Machine Learning Research}, 21\penalty0
  (54):\penalty0 1--5, 2020.

\bibitem[G{\"o}nen and Alpayd{\i}n(2011)]{gonen2011multiple}
Mehmet G{\"o}nen and Ethem Alpayd{\i}n.
\newblock Multiple kernel learning algorithms.
\newblock \emph{The Journal of Machine Learning Research}, 12:\penalty0
  2211--2268, 2011.

\bibitem[{Monroe} et~al.(2021){Monroe}, {Campbell}, {Duan}, {Gong}, {Gorshkov},
  {Hess}, {Islam}, {Kim}, {Linke}, {Pagano}, {Richerme}, {Senko}, and
  {Yao}]{Monroe21}
C.~{Monroe}, W.~C. {Campbell}, L.~M. {Duan}, Z.~X. {Gong}, A.~V. {Gorshkov},
  P.~W. {Hess}, R.~{Islam}, K.~{Kim}, N.~M. {Linke}, G.~{Pagano},
  P.~{Richerme}, C.~{Senko}, and N.~Y. {Yao}.
\newblock {Programmable quantum simulations of spin systems with trapped ions}.
\newblock \emph{Reviews of Modern Physics}, 93\penalty0 (2):\penalty0 025001,
  April 2021.
\newblock \doi{10.1103/RevModPhys.93.025001}.

\bibitem[Berloff et~al.(2017)Berloff, Kalinin, Silva, Langbein, and
  Lagoudakis]{berloff2017}
Natalia~G Berloff, Kirill Kalinin, Matteo Silva, Wolfgang Langbein, and
  Pavlos~G Lagoudakis.
\newblock Realizing the xy hamiltonian in polariton simulators.
\newblock \emph{Nature Materials}, 16:\penalty0 1120–1126, 2017.
\newblock URL \url{https://doi.org/10.1038/nmat4971}.

\bibitem[Browaeys and Lahaye(2020)]{browaeys2020many}
Antoine Browaeys and Thierry Lahaye.
\newblock Many-body physics with individually controlled rydberg atoms.
\newblock \emph{Nature Physics}, 16\penalty0 (2):\penalty0 132--142, Feb 2020.
\newblock ISSN 1745-2481.
\newblock \doi{10.1038/s41567-019-0733-z}.
\newblock URL \url{https://doi.org/10.1038/s41567-019-0733-z}.

\bibitem[{Saffman} et~al.(2010){Saffman}, {Walker}, and {M{\o}lmer}]{Saffman10}
M.~{Saffman}, T.~G. {Walker}, and K.~{M{\o}lmer}.
\newblock {Quantum information with Rydberg atoms}.
\newblock \emph{Reviews of Modern Physics}, 82\penalty0 (3):\penalty0
  2313--2363, Jul 2010.
\newblock \doi{10.1103/RevModPhys.82.2313}.

\bibitem[{Saffman}(2016)]{Saffman2016}
M.~{Saffman}.
\newblock {Quantum computing with atomic qubits and Rydberg interactions:
  progress and challenges}.
\newblock \emph{Journal of Physics B Atomic Molecular Physics}, 49\penalty0
  (20):\penalty0 202001, October 2016.
\newblock \doi{10.1088/0953-4075/49/20/202001}.

\bibitem[Barredo et~al.(2016)Barredo, L{\'e}s{\'e}leuc, Lienhard, Lahaye, and
  Browaeys]{barredo_atom-by-atom_2016}
Daniel Barredo, Sylvain~de L{\'e}s{\'e}leuc, Vincent Lienhard, Thierry Lahaye,
  and Antoine Browaeys.
\newblock An atom-by-atom assembler of defect-free arbitrary two-dimensional
  atomic arrays.
\newblock \emph{Science}, 354\penalty0 (6315):\penalty0 1021--1023, November
  2016.
\newblock ISSN 0036-8075, 1095-9203.
\newblock \doi{10.1126/science.aah3778}.
\newblock URL \url{https://science.sciencemag.org/content/354/6315/1021}.

\bibitem[Endres et~al.(2016)Endres, Bernien, Keesling, Levine, Anschuetz,
  Krajenbrink, Senko, Vuletic, Greiner, and Lukin]{endres_atom-by-atom_2016}
Manuel Endres, Hannes Bernien, Alexander Keesling, Harry Levine, Eric~R.
  Anschuetz, Alexandre Krajenbrink, Crystal Senko, Vladan Vuletic, Markus
  Greiner, and Mikhail~D. Lukin.
\newblock Atom-by-atom assembly of defect-free one-dimensional cold atom
  arrays.
\newblock \emph{Science}, 354\penalty0 (6315):\penalty0 1024--1027, November
  2016.
\newblock ISSN 0036-8075, 1095-9203.
\newblock \doi{10.1126/science.aah3752}.
\newblock URL \url{https://science.sciencemag.org/content/354/6315/1024}.

\bibitem[Barredo et~al.(2018)Barredo, Lienhard, L{\'e}s{\'e}leuc, Lahaye, and
  Browaeys]{barredo_synthetic_2018}
Daniel Barredo, Vincent Lienhard, Sylvain~de L{\'e}s{\'e}leuc, Thierry Lahaye,
  and Antoine Browaeys.
\newblock Synthetic three-dimensional atomic structures assembled atom by atom.
\newblock \emph{Nature}, 561\penalty0 (7721):\penalty0 79--82, September 2018.
\newblock ISSN 1476-4687.
\newblock \doi{10.1038/s41586-018-0450-2}.
\newblock URL \url{https://www.nature.com/articles/s41586-018-0450-2}.

\bibitem[Henriet et~al.(2020)Henriet, Beguin, Signoles, Lahaye, Browaeys,
  Reymond, and Jurczak]{Henriet2020quantum}
Lo{\"{i}}c Henriet, Lucas Beguin, Adrien Signoles, Thierry Lahaye, Antoine
  Browaeys, Georges-Olivier Reymond, and Christophe Jurczak.
\newblock Quantum computing with neutral atoms.
\newblock \emph{{Quantum}}, 4:\penalty0 327, September 2020.
\newblock ISSN 2521-327X.
\newblock \doi{10.22331/q-2020-09-21-327}.
\newblock URL \url{https://doi.org/10.22331/q-2020-09-21-327}.

\bibitem[{Morgado} and {Whitlock}(2020)]{Morgado20}
M.~{Morgado} and S.~{Whitlock}.
\newblock {Quantum simulation and computing with Rydberg-interacting qubits}.
\newblock \emph{arXiv e-prints}, art. arXiv:2011.03031, November 2020.

\bibitem[{Beterov}(2020)]{Beterov20}
I.~I. {Beterov}.
\newblock {Quantum Computers Based on Cold Atoms}.
\newblock \emph{Optoelectronics, Instrumentation and Data Processing},
  56\penalty0 (4):\penalty0 317--324, July 2020.
\newblock \doi{10.3103/S8756699020040020}.

\bibitem[{Wu} et~al.(2020){Wu}, {Liang}, {Tian}, {Yang}, {Chen}, {Liu}, {Khoon
  Tey}, and {You}]{Wu21}
Xiaoling {Wu}, Xinhui {Liang}, Yaoqi {Tian}, Fan {Yang}, Cheng {Chen},
  Yong-Chun {Liu}, Meng {Khoon Tey}, and Li~{You}.
\newblock {A concise review of Rydberg atom based quantum computation and
  quantum simulation}.
\newblock \emph{arXiv e-prints}, art. arXiv:2012.10614, December 2020.

\bibitem[Schau{\ss} et~al.(2015)Schau{\ss}, Zeiher, Fukuhara, Hild, Cheneau,
  Macr{\`\i}, Pohl, Bloch, and Gro{\ss}]{schauss2015crystallization}
Peter Schau{\ss}, Johannes Zeiher, Takeshi Fukuhara, Sebastian Hild, Marc
  Cheneau, Tommaso Macr{\`\i}, Thomas Pohl, Immanuel Bloch, and Christian
  Gro{\ss}.
\newblock Crystallization in ising quantum magnets.
\newblock \emph{Science}, 347\penalty0 (6229):\penalty0 1455--1458, 2015.

\bibitem[Labuhn et~al.(2016)Labuhn, Barredo, Ravets, De~L{\'e}s{\'e}leuc,
  Macr{\`\i}, Lahaye, and Browaeys]{labuhn2016tunable}
Henning Labuhn, Daniel Barredo, Sylvain Ravets, Sylvain De~L{\'e}s{\'e}leuc,
  Tommaso Macr{\`\i}, Thierry Lahaye, and Antoine Browaeys.
\newblock Tunable two-dimensional arrays of single rydberg atoms for realizing
  quantum ising models.
\newblock \emph{Nature}, 534\penalty0 (7609):\penalty0 667--670, 2016.

\bibitem[de~L\'es\'eleuc et~al.(2018)de~L\'es\'eleuc, Weber, Lienhard, Barredo,
  B\"uchler, Lahaye, and Browaeys]{leseleuc2018accurate}
Sylvain de~L\'es\'eleuc, Sebastian Weber, Vincent Lienhard, Daniel Barredo,
  Hans~Peter B\"uchler, Thierry Lahaye, and Antoine Browaeys.
\newblock Accurate mapping of multilevel rydberg atoms on interacting
  spin-$1/2$ particles for the quantum simulation of ising models.
\newblock \emph{Phys. Rev. Lett.}, 120:\penalty0 113602, Mar 2018.
\newblock \doi{10.1103/PhysRevLett.120.113602}.
\newblock URL \url{https://link.aps.org/doi/10.1103/PhysRevLett.120.113602}.

\bibitem[Barredo et~al.(2015)Barredo, Labuhn, Ravets, Lahaye, Browaeys, and
  Adams]{barredo2015coherent}
Daniel Barredo, Henning Labuhn, Sylvain Ravets, Thierry Lahaye, Antoine
  Browaeys, and Charles~S. Adams.
\newblock Coherent excitation transfer in a spin chain of three rydberg atoms.
\newblock \emph{Phys. Rev. Lett.}, 114:\penalty0 113002, Mar 2015.
\newblock \doi{10.1103/PhysRevLett.114.113002}.
\newblock URL \url{https://link.aps.org/doi/10.1103/PhysRevLett.114.113002}.

\bibitem[Orioli et~al.(2018)Orioli, Signoles, Wildhagen, G\"unter, Berges,
  Whitlock, and Weidem\"uller]{orioli2018relaxation}
A.~Pi\~neiro Orioli, A.~Signoles, H.~Wildhagen, G.~G\"unter, J.~Berges,
  S.~Whitlock, and M.~Weidem\"uller.
\newblock Relaxation of an isolated dipolar-interacting rydberg quantum spin
  system.
\newblock \emph{Phys. Rev. Lett.}, 120:\penalty0 063601, Feb 2018.
\newblock \doi{10.1103/PhysRevLett.120.063601}.
\newblock URL \url{https://link.aps.org/doi/10.1103/PhysRevLett.120.063601}.

\bibitem[de~L{\'e}s{\'e}leuc et~al.(2019)de~L{\'e}s{\'e}leuc, Lienhard, Scholl,
  Barredo, Weber, Lang, B{\"u}chler, Lahaye, and Browaeys]{deleseleuc19}
Sylvain de~L{\'e}s{\'e}leuc, Vincent Lienhard, Pascal Scholl, Daniel Barredo,
  Sebastian Weber, Nicolai Lang, Hans~Peter B{\"u}chler, Thierry Lahaye, and
  Antoine Browaeys.
\newblock Observation of a symmetry-protected topological phase of interacting
  bosons with rydberg atoms.
\newblock \emph{Science}, 365\penalty0 (6455):\penalty0 775--780, 2019.
\newblock ISSN 0036-8075.
\newblock \doi{10.1126/science.aav9105}.
\newblock URL \url{https://science.sciencemag.org/content/365/6455/775}.

\bibitem[Silvério et~al.(2021)Silvério, Grijalva, Dalyac, Leclerc, Karalekas,
  Shammah, Beji, Henry, and Henriet]{Pulser}
Henrique Silvério, Sebastián Grijalva, Constantin Dalyac, Lucas Leclerc,
  Peter~J. Karalekas, Nathan Shammah, Mourad Beji, Louis-Paul Henry, and Loïc
  Henriet.
\newblock Pulser: An open-source package for the design of pulse sequences in
  programmable neutral-atom arrays, 2021.

\bibitem[{Tang}(2018)]{Tang18}
Ewin {Tang}.
\newblock {Quantum-inspired classical algorithms for principal component
  analysis and supervised clustering}.
\newblock \emph{arXiv e-prints}, art. arXiv:1811.00414, October 2018.

\bibitem[Arrazola et~al.(2020)Arrazola, Delgado, Bardhan, and
  Lloyd]{Arrazola20}
Juan~Miguel Arrazola, Alain Delgado, Bhaskar~Roy Bardhan, and Seth Lloyd.
\newblock Quantum-inspired algorithms in practice.
\newblock \emph{{Quantum}}, 4:\penalty0 307, August 2020.
\newblock ISSN 2521-327X.
\newblock \doi{10.22331/q-2020-08-13-307}.
\newblock URL \url{https://doi.org/10.22331/q-2020-08-13-307}.

\bibitem[Frazier(2018)]{frazier2018tutorial}
Peter~I. Frazier.
\newblock A tutorial on bayesian optimization, 2018.

\bibitem[Matérn(1986)]{Matern60}
Bertil Matérn.
\newblock \emph{{Spatial Variation}}.
\newblock {Springer-Verlag}, 1986.

\bibitem[Kandasamy et~al.(2018)Kandasamy, Krishnamurthy, Schneider, and
  P{\'o}czos]{kandasamy2018parallelised}
Kirthevasan Kandasamy, Akshay Krishnamurthy, Jeff Schneider, and Barnab{\'a}s
  P{\'o}czos.
\newblock Parallelised bayesian optimisation via thompson sampling.
\newblock In \emph{International Conference on Artificial Intelligence and
  Statistics}, pages 133--142. PMLR, 2018.

\end{thebibliography}

\end{document}